\documentclass[12pt]{article}
\usepackage[nohead, margin = 1in, footskip = .5cm]{geometry}
\usepackage{rotating}
\usepackage{longtable}
\usepackage{hyperref}
\usepackage{booktabs}
\usepackage{dcolumn}
\usepackage[flushleft]{threeparttable}
\usepackage{rotating}
\usepackage{setspace}
\usepackage{times}
\usepackage{tabularx}
\hypersetup{colorlinks, linkcolor = blue, citecolor = blue, urlcolor = blue}
\usepackage{natbib}
\usepackage{amsmath}
\usepackage{graphicx}
\usepackage{float}
\floatstyle{plaintop}
\restylefloat{table}
\restylefloat{figure}
\usepackage[tableposition=top, figureposition=top, labelfont=bf, font = normal]{caption}
\usepackage{array}
\usepackage{subcaption}
\usepackage{siunitx}
\usepackage[table]{xcolor}
\usepackage{multirow}
\usepackage{hhline}
\usepackage{calc}
\usepackage{tabularx}
\usepackage[nameinlink]{cleveref}
\usepackage{bm}
\usepackage{caption}
\usepackage{amssymb}

\newcommand{\ratefirstevertreat}{78.4}
\newcommand{\ratefirsteverctr}{72.8}
\newcommand{\ratefirsteverdiff}{5.6}

\newcommand{\ratethirdeverdiff}{1.7}

\newcommand{\tightnessexp}{31.6}
\newcommand{\tightnessall}{18.1}

\newcommand{\minsbtwndiff}{34}
\newcommand{\sharepreexp}{45}
\newcommand{\sharemissingemail}{2.2}

\newcommand{\diffpredictonerev}{17}
\newcommand{\diffpredictonenights}{0.2}
\newcommand{\diffpredictfrev}{8}
\newcommand{\diffpredictfnights}{0.1}
\newcommand{\diffpredicttfrev}{4}
\newcommand{\diffpredicttfnights}{0.05}
\newcommand{\regestimatenights}{0.02}
\newcommand{\regestimaterev}{4.26}

\newcommand{\nctr}{326,515}
\newcommand{\ntrt}{328,080}
\newcommand{\reviewrateeffect}{12.85}
\newcommand{\reviewratectr}{24.17}
\newcommand{\reviewratetrt}{37.01}
\newcommand{\reviewrateratio}{53}
\newcommand{\reviewrateone}{0.2}
\newcommand{\reviewratetwo}{0.31}
\newcommand{\reviewratethree}{1.44}
\newcommand{\reviewratefour}{4.64}
\newcommand{\reviewratefive}{6.26}
\newcommand{\condavgrate}{0.07}
\newcommand{\condavgtrt}{4.41}
\newcommand{\condavgctr}{4.48}

\newcommand{\meanposctr}{94.1}
\newcommand{\meanpostrt}{93.2}

\newcommand{\perlendif}{8}

\newcommand{\effecatelateview}{7.1}
\newcommand{\effecatelatetrans}{9.1}
\newcommand{\effecatelatetranslevel}{0.326}

\begin{document}
\title{More Reviews May Not Help: Evidence from Incentivized First Reviews on Airbnb} 
\author{Andrey Fradkin\thanks{fradkin@bu.edu, Primary Author}  \, David Holtz\thanks{dholtz@haas.berkeley.edu}\thanks{We thank Dean Eckles, Chiara Farronato, Shane Greenstein, John Horton, Caroline Hoxby, Xiang Hui, Ramesh Johari, Garrett Johnson, Jon Levin, Tesary Lin, Mike Luca, Steve Tadelis, Catherine Tucker, Giorgos Zervas, and seminar participants at Microsoft, ACM EC'15, NBER Summer Institute, CODE, WISE, Marketing Science (2021), and the University of Rochester for comments. We thank Elena Grewal and Riley Newman for giving us the initial opportunity to work on this project, Matthew Pearson for early conversations about the project, and Peter Coles and Mike Egesdal for their tireless efforts in helping this paper be approved. The views expressed in this paper are solely the authors' and do not necessarily reflect the views of Airbnb, Inc. The authors were employed by Airbnb, Inc. for part of the time that this paper was written and have held stock that may constitute a material financial position.}}
\maketitle

\begin{abstract}
	Online reviews are typically written by volunteers and, as a consequence, information about seller quality may be under-provided in digital marketplaces. We study the extent of this under-provision in a large-scale randomized experiment conducted by Airbnb. In this experiment, buyers are offered a coupon to review listings that have no prior reviews. The treatment induces additional reviews and these reviews tend to be more negative than reviews in the control group, consistent with selection bias in reviewing. Reviews induced by the treatment result in a temporary increase in transactions but these transactions are for fewer nights, on average. The effects on transactions and nights per transaction cancel out so that there is no detectable effect on total nights sold and revenue. Measures of transaction quality in the treatment group fall, suggesting that incentivized reviews do not improve matching. We show how market conditions and the design of the reputation system can explain our findings.
\end{abstract}
\clearpage
\doublespacing

\section{Introduction}

Reputation systems are used by nearly every digital marketplace to help match buyers and sellers. Although reputation systems are considered critical to the success of digital marketplaces, they are known to suffer from a variety of biases and imperfections \citep{tadelis2016reputation}. Bias in reviews arises when the distribution of observed reviews does not accurately represent the distribution of actual transaction quality. We study the consequences of review bias on market outcomes through an experimental evaluation of an incentivized review policy in which buyers are provided with an incentive to review sellers.

Incentives to review have the potential to address bias arising from non-random selection into which transactions are reviewed. Prior work starting with \citet{Dellarocas.2007} has identified that those with positive or more extreme experiences review at rates above average and that incentives to review reduce the extent of this selection based on quality (\citet{burtch2018stimulating, marinescu2021incentives}). However, this work has not measured whether reductions in review bias from incentives actually improve market outcomes. Mapping review bias to market outcomes is critical since the purpose of reviews is to create better matches and since incentives should pass a cost-benefit analysis prior to adoption.

There are two theoretical reasons by which the under-provision of accurate reviews may harm market outcomes and by which incentives to review can help. First, incentives to review can increase the speed of learning about seller quality by generating more reviews. \citet{acemoglu2019learning} show that faster learning about seller quality increases welfare. Second, biased reviews may result in the formation of bad matches because buyers may be less aware of negative or polarizing characteristics of a seller. Reviews induced by incentives are more likely to generate information about these characteristics. The above reasons may be especially relevant for sellers without prior reviews, since there is a lot of uncertainty about the quality of services they provide (\citet{Pallais.2014}). 

We conduct the first field experimental analysis of an incentivized review policy that also measures the effects of incentivized reviews on subsequent market outcomes. We find that, although incentives increase review rates and lower average ratings, they do not generate economically meaningful benefits to the platform or its users. The experiment was conducted by Airbnb between 2014 and 2016 and provided a \$25 coupon in exchange for a review of a listing without reviews. In particular, guests to treated listings were sent an email offering them an Airbnb coupon in exchange for a review if they had not reviewed within (typically) 8 or 9 days after checkout, while guests to control listings were instead sent an email reminding them to leave a review.

The incentivized first review policy increased the review rate by \reviewrateratio{}\% and resulted in reviews with a lower average rating. These reviews caused a transitory increase in transactions for treated sellers. In particular, an induced review increased transactions by \effecatelatetrans{}\% in the 120 days following treatment assignment. However, these transactions were for fewer nights, and as a consequence the revenue effects of these reviews were smaller and not statistically distinguishable from zero. Consequently neither treated sellers nor the platform, which charges a percentage fee, benefitted from the policy.

We also show that the induced reviews do not improve transaction quality as measured by the customer complaint rates, subsequent reviews, and the post-transaction usage of Airbnb by guests. This finding, along with the finding about the lack of revenue effects, suggests that reviews were not underprovided for the subset of transaction for which reviews were induced. Because the policy had costs and resulted in small measurable benefits, the policy was not beneficial to the platform or its users. 

Our results may appear puzzling given that the prior literature has found that reviews can have large causal effects on demand. We argue that these differences arise because we are interested in a different causal estimand and because of the institutional context of Airbnb. In particular, we study the short and long-run effects of a first review for a set of reviews induced by our policy. This is the policy relevant parameter for evaluating incentivized review policies. Other papers focus on the effect of a positive review relative to a negative review under a particular selection of reviews (\citet{park2021fateful}), the effect of average star ratings (\citet{lewis2016welfare}), or the effect of hiring and reviewing a worker without experience (\citet{Pallais.2014}). 

Furthermore, the market structure of Airbnb has an influence on the effects of incentivized reviews. Most listings in our sample would have been reviewed eventually, regardless of treatment status. This is because listings often have other transactions in progress and these transactions also result in reviews. Reviews from other transactions arrive quickly --- the difference in the median time of first reviews between the treatment and control group is just 6 days. The fact that there is a lag between a booking and a review is not unique to Airbnb. Similar dynamics occur in peer-to-peer markets such as Etsy, Rover, Uber, and Upwork, although the speed of reviews varies across marketplaces. In contrast, the work of \citet{Pallais.2014} focuses on solving the `cold-start' problem for sellers who cannot obtain transactions without reviews.

A second reason why incentivized reviews may have small effects on listing outcomes is due to the way in which Airbnb's review system displayed text and ratings during the time of the experiment (2014 to 2016). In particular, star ratings (as opposed to the text and number of reviews) were only displayed once a listing had at least three ratings and as an average rounded to the nearest half a point. As a result, an induced rating was averaged with at least two other ratings when displayed to guests on Airbnb. This averaging and rounding attenuated perceived differences between the ratings of control and treatment listings.\footnotemark{} For text, which is always displayed, we find that treated reviews have a one percentage point higher rate of text classified as negative compared to the control group.  

\footnotetext{\citet{park2021fateful} find large negative effects of a first review with a low rating for vacuum cleaners and toasters. In our setting, a first review with a low star rating is likely to have smaller effects since it would be averaged with at least two other ratings and rounded prior to being shown. Rounding is also used by Yelp, Etsy, and Facebook Marketplace.}

An implication of our findings is that institutional details such as market conditions and reputation system design are critical for understanding the role of reviews and the effect of reputation system designs. We do not claim that that reviews and review systems have little value. Indeed, prior work has shown that reputation systems substantially increase consumer surplus (\citet{lewis2016welfare}, \citet{reimers2021digitization}, and \citet{wu2015economic}). Instead, we show that additional reviews do not matter when listings are expected to receive a flow of reviews and when review ratings are displayed as averages. The incentivized review policy we study also had imperfect compliance, \reviewratetrt{}\% of treated transactions resulted in a review. A policy that induced reviews for a different subset of transactions could have different effects on market outcomes.

The rest of the paper proceeds as follows. We provide an overview of the related research literature in \autoref{sec:litreview} and  a theoretical framework for understanding the effects of induced reviews in \autoref{sec:theory}. Next, we describe the experiment in \autoref{sec:expdesign} and the treatment's effects on reviews in \autoref{sec:effectsonreviews}. Lastly, \autoref{sec:effectonoutcomes} empirically studies the implication of incentivized reviews for market outcomes and \autoref{sec:conclusion} concludes.

\section{Literature Review} \label{sec:litreview} 

We contribute to four related research themes within the study of online marketplaces. The first research theme studies the impacts of incentivized review policies and nudges to review in online markets. The second research theme concerns review biases and `fake` reviews. The third research theme focuses on field experiments that attempt to measure the impact of different reputation system designs, and the fourth research theme concerns models of learning based on consumer reviews.

Because evaluations will be underprovided in equilibrium in the absence of an appropriate payment scheme \citep{avery1999market}, a number of recent papers have studied the effectiveness of incentivized review policies and nudges to review at increasing review rates in online settings. \citet{burtch2018stimulating},  \citet{marinescu2021incentives}, and \citet{karaman2020online} document that monetary incentives, non-monetary incentives, and review solicitation emails all reduce review bias caused by selection and bring the observed distribution of ratings closer to the distribution of experiences that buyers have. While each of these studies documents the impacts of platform-driven interventions on the accuracy of consumer feedback, they do not study the downstream impacts of the induced reviews on market outcomes. In a related stream of work, \citet{li2010reputation}, \citet{li2014money}, \citet{cabral2015dollar}, and \citet{li2020buying} study policies in which the seller (rather than the platform) offers a rebate for a review. The theoretical basis for these policies is that when sellers have the option to provide rebates, review rates increase, ratings are less biased, and there is less moral hazard \citep{li2010reputation}. The rebate for review mechanisms proposed in these papers have two important differences from the policy studied in this work. First, sellers select into offering a rebate, meaning that the rebate program is not universal and the rebate offer may serve as a signal of quality (as shown in \citet{li2020buying} for sellers on Taobao). Second, because the rebate offer comes from the seller, the buyer may reciprocate the seller's rebate by biasing their review upwards \citep{cabral2015dollar}. 

Beyond work that studies incentivized review policies and nudges to review, there is also a broader research literature that quantifies the ways in which online reviews can misrepresent the experiences of buyers. Several papers show that non-incentivized reviews exhibit a positivity bias \citep{Dellarocas.2007, nosko2015limits, filippas2018reputation, brandes2019drives, fradkin2020reciprocity}. This positivity bias has a number of causes, including a higher propensity to post feedback when satisfied \citep{Dellarocas.2007}, reciprocity and retaliation \citep{Dellarocas.2007, fradkin2020reciprocity}, selection effects that dictate which buyers transact with which sellers \citep{nosko2015limits}, `reputation inflation' \citep{filippas2018reputation}, and differential attrition of those with moderate experiences \citep{brandes2019drives}. \citet{hui2021and} show that the decision to review depends on the change in buyers beliefs about seller quality before and after a transaction.\footnotemark{}

\footnotetext{A related literature studies `fake' reviews \citep{luca2016fake, mayzlin2014promotional, he2020market}, however, fake reviews are less of a concern in our research setting because all Airbnb reviews are linked to a particular transaction.}

There is also a growing research literature that uses field experiments to study the impacts of changes to reputation system design on subsequent market outcomes, rather than on just the types of reviews that are left \citep{cabral2015dollar, benson2020can, fradkin2020reciprocity, hui2020mitigating, garg2021designing, Pallais.2014}.\footnotemark{} One existing piece of research that is particularly closely related to our work is \citet{Pallais.2014}, which experimentally measures the effects of an intervention in which new sellers are both hired and reviewed. Pallais finds that hiring workers and leaving positive feedback has large positive effects on subsequent demand. In contrast to \citet{Pallais.2014}, the policy we study generates reviews only for the subset of sellers who are able to transact before receiving their first review. An advantage of this policy relative to the one studied in \citet{Pallais.2014} is that it is less expensive for platforms to implement, since they do not have to pay to hire new sellers. However, a disadvantage is that the incentivized first review policy does not solve the `cold-start' problem for listings who are unable to transact prior to receiving their first review.

\footnotetext{A preliminary analysis of this experiment was presented in \citet{fradkin2015bias}. This analysis focused on the first month of data and did not study market outcomes.}

Finally, there is an emerging research literature that aims to model the dynamics with which consumers learn about seller quality under different ratings systems \citep{papanastasiou2018crowdsourcing, besbes2018information, acemoglu2019learning, ifrach2019bayesian}. Our work is particularly closely related to \citet{besbes2018information} and \citet{acemoglu2019learning}, which both compare consumer learning dynamics under reputation systems in which the full rating history is displayed to learning dynamics under ratings systems in which only summary statistics are displayed. In addition to the fact that the theoretical framework we use to interpret our results is a simplified version of the model presented in \citet{acemoglu2019learning}, we will argue in this paper that one of reasons that the additional reviews induced by Airbnb's incentivized first review program failed to have a measurable impact on market outcomes is the design of Airbnb's review system, which has characteristics of the ``summary statistics only'' systems discussed in both \citet{acemoglu2019learning} and \citet{besbes2018information}.

We contribute to these research literatures by experimentally studying the implications of a change to the design of Airbnb's reputation system not only on reviews, but also on subsequent market outcomes.\footnotemark{} Although one might expect an incentivized first review program like the one we study to yield additional information on seller quality that yields more and better matches on the platform, we do not find this to be the case empirically; we find that although the incentivized review program reduced bias in the reviews left by Airbnb guests, they had negligible effects on demand and match quality. We argue that this is due to market conditions in the markets where most of our sample listings were located, as well as the manner in which Airbnb's reputation system aggregates and displays review information.

\footnotetext{\citet{laouenan2020can} and \citet{cui2020reducing} study the effects of Airbnb reviews on market outcomes with a focus on discrimination.}

\section{Theoretical Framework}\label{sec:theory}

Whether the platform should incentivize reviews depends on whether the induced reviews improve outcomes on the platform. In this section, we describe a theoretical framework which clarifies the conditions under which incentivized reviews increase demand and the utility of buyers. Our framework is is a simplified version of \citet{acemoglu2019learning}, which characterizes the speed of learning in review systems and shows that review systems with a higher speed of learning increase expected the expected utility of buyers. We derive a formal model of this process in \autoref{app:theory}. 

In our theoretical framework, the degree to which incentivized reviews improve the utility of subsequent buyers is a function of the informativeness of the review system. We conceptualize the informativeness of a review system by the extent to which buyer beliefs about quality after seeing review information (or lack thereof) correspond to the true quality of a listing. This informativeness is a function of both the extent to which ratings correlate with quality and the extent to which buyers' beliefs about reviews correspond to rational expectations. Note that horizontal preferences across listings can be accommodated if buyers first condition on characteristics such as the listing description and photos. 

Suppose that a listing has one of two qualities, good or bad, and one of three post-transaction review outcomes: a negative review, no review, and a positive review. Our treatment induces reviews for listings who would have no review were they in the control group.\footnotemark{} The effects of the treatment on demand are a function of two terms. The first term represents the share of treated listings for which the treatment changes the review outcome from no review to positive review. This is multiplied by the effect on individual demand of one positive review. The demand effect of a positive review corresponds to the change in a buyer's belief that a listing is high quality when a positive review is present. 

\footnotetext{For the purposes of the theoretical framework, we also assume that the coupon offer does not change the reviews of those guests who would have reviewed regardless of the coupon.}

Similarly, there is a term that represents the share of listings for whom the induced review is negative times the negative effect on demand of a negative review (vs no review). The average treatment effect is the sum of these two terms, which could be positive or negative. Our analysis in \autoref{sec:effectsonreviews} shows that the there is a much bigger increase in positive reviews than in negative reviews. However, it is possible that the increase in demand due to the positive reviews is small and the decrease in demand due to negative reviews is large, in which case the effect of incentivized reviews may be negative. 

The strength of any demand effects of one review depend on the extent to which buyer beliefs are updated. Bayesian updating suggests that buyer beliefs about quality should change more when no other reviews are present than when other reviews are present. As a result, the effects of incentivized reviews are mediated by whether the listing is able to quickly obtain other reviews. We document the presence of reviews from non-focal transactions in \autoref{sec:whysmalleffect}.

The effects of incentivized reviews on expected utility in the model are more subtle than the effects on demand. If reviews always corresponded to quality, then incentivized reviews would help buyers identify good and bad listings more quickly. This would increase consumer utility. However, since reviews do not perfectly correlate with quality, low quality listings may be reviewed positively and high quality listings may be reviewed negatively. If incentives cause enough low quality listings to be reviewed positively or enough high quality listings to be reviewed negatively, then the utility of consumers may actually fall due to incentivized reviews. Whether this happens in practice depends on the composition (high or low quality) of non-reviewed listings for whom the incentive induces a review. The net sum of all of these terms is an empirical question which we explore in the next section.

Lastly, there are two additional considerations in our empirical investigation that fall outside of the simple framework above. First, reviews may help in matching not just vertically but also horizontally. For example, a review could increase the salience of some characteristics, such as the presence of pets, which some guests  like and others do not. To study whether reviews affect horizontal matching, we consider whether the characteristics of guests or trips change due to our treatment. Secondly, reviews may change the behavior of sellers, as suggested in \citet{hui2018adverse} and \citet{klein2016market}. We investigate this margin by studying differences between the treatment and control groups in how hosts portray their properties through photos and descriptions. 

\section{Setting and Experimental Design}\label{sec:expdesign}

We analyze an experiment conducted on Airbnb, the largest online marketplace for peer-to-peer short-term accommodations, from April 12, 2014 to May 17, 2016. At the time of the experiment, Airbnb's review system worked as follows. After the guest's checkout, both hosts and guests were asked via email, web notifications, and app notifications to review each other. Both guest and host reviews consisted of both numeric and textual information. The text of reviews written by guests were displayed on listing pages in reverse chronological order. The numeric overall and category ratings,\footnotemark{}\footnotetext{The six categories were: the accuracy of the listing compared to the guest’s expectations, the communication of the host, the cleanliness of the listing, the location of the listing, the value of the listing, and the quality of the amenities provided by the listing.} which were on a one- to five-star scale, were displayed as averages across all transactions rounded to the nearest half star.\footnotemark{}\footnotetext{Airbnb has, subsequent to the period covered in our analysis sample, switched to displaying star rating averages rounded to the nearest second decimal point.} Rounded average ratings were only visible on listing pages once a lasting had received at least three reviews; before that, only review text was visible on a listing page. The number of reviews was visible on both the search and listing pages as long as the listing had one review, implying that reviews can have effects both through the search page and through the listing page. 

Prior to July 2014, both guest and host reviews were visible both to the counterparty and to the public immediately after submission, and reviews needed to be submitted within 30 days of checkout. Beginning in July 2014, a simultaneous reveal review system was in place.\footnotemark{}\footnotetext{The simultaneous reveal review system was evaluated using a platform-wide A/B test that ran from May 2014 to July 2014. See \citet{fradkin2020reciprocity} for details.} Under the simultaneous reveal system, guests and hosts had 14 days after checkout to submit a review, and reviews were only publicly displayed after both parties submitted a review or 14 days had elapsed. Because our experiment ran from April 2014 to May 2016, the vast majority of our data was collected under the simultaneous reveal review system.

Experiment randomization was conducted at the Airbnb listing level. In order to be eligible for enrollment in the experiment, a listing needed to meet the following criteria:

\begin{itemize}
	\item It needed to have been booked.
	\item It needed to have no prior reviews.
	\item Following a guest checkout, the guest must not have reviewed within a threshold number of days. This number was typically 8 or 9 days throughout most of our sample, with the specific number of days being a function of the email dispatching system.\footnotemark{} 
\end{itemize}

\footnotetext{The time between review eligibility and an email from the experiment was influenced by the day of the week and time of the year. After March of 2016, the number of days within which a review must have been submitted was changed to 7. See \autoref{app:daystoemail} for an a more detailed discussion.}

\begin{figure}
	\centering
	\includegraphics[width=.8\textwidth]{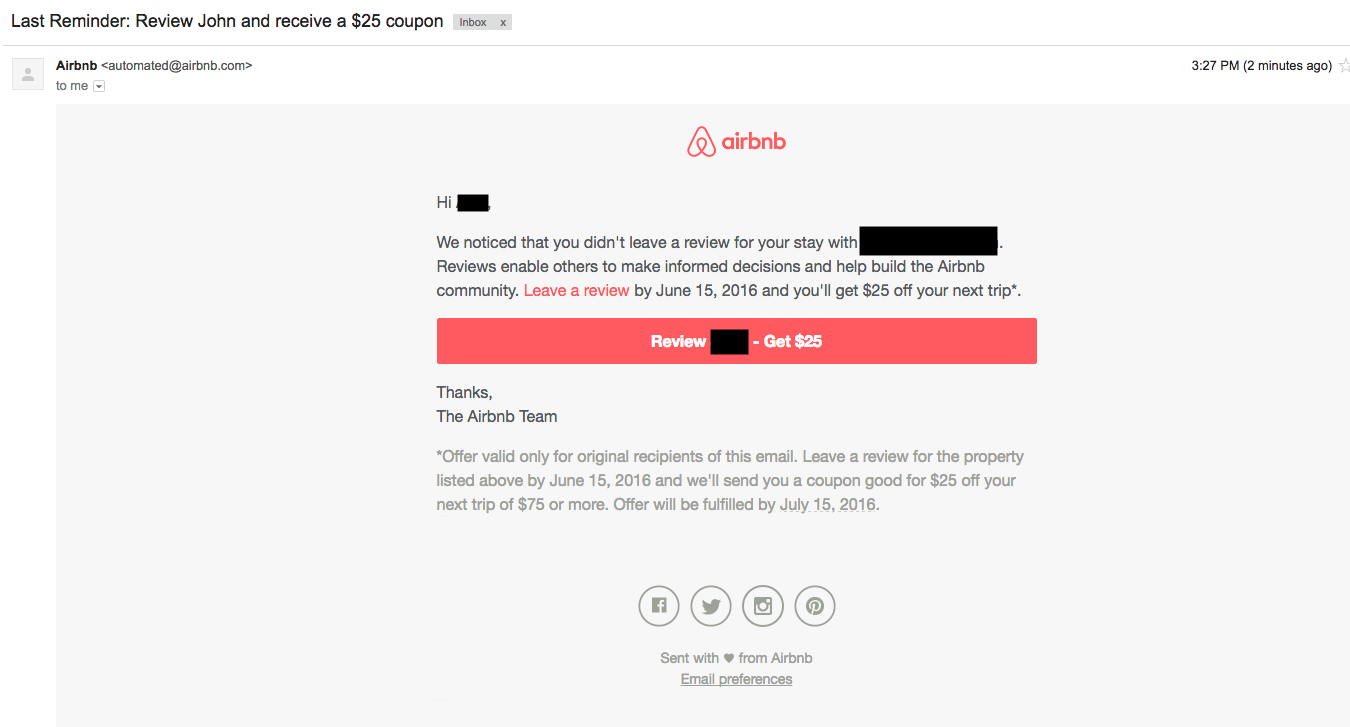}
	\caption{Treatment Email}
		\label{fig:incent_email_treatment}
	\footnotesize \emph{Notes:} \raggedright  Displays the email sent to guests who had stayed in treatment listings who had not yet received a review on Airbnb after a certain number of days, inviting them to leave a review in exchange for a coupon. 
\end{figure}

Across Airbnb's entire platform, guests who had not reviewed within the threshold number of days described above received an email reminding them to review. For stays at listings that met the criteria above and were assigned to the control group, guests received the standard review reminder email. For stays at listings that met the criteria above and were assigned to the treatment group, guests received a reminder email that also offered a \$25 Airbnb coupon in exchange for leaving a review. These coupons expired one year after being issued, and needed to be used on stays with a minimum cost of \$75. Figure \ref{fig:incent_email_treatment} shows the email sent to guests who stayed at treatment listings without reviews during the experiment. In our sample, \nctr{} listings were assigned to the control, whereas \ntrt{} listings were assigned to the treatment. 
The experiment achieved good balance on pre-treatment covariates and used a well-tested system at Airbnb (\autoref{fig:balance}).

\section{Effects of Experiment on Reviews}\label{sec:effectsonreviews}

In this section, we show that the coupon email treatment induces reviews and that these reviews tend to be worse on average. We first measure the effect of the treatment on the types of ratings and reviews left by guests in either treatment arm who had stays meeting the criteria required to receive the incentive in the treatment group. In particular, we call the first transaction for which a listing is either in the treatment or control the \textit{focal stay}, in contrast to subsequent stays that may also have resulted in reviews. We show that the treatment induced reviews for the focal stay and that those reviews tended to have lower ratings on average. While the textual content of those reviews also tended to contain more negative sentiment on average, there was no difference in text sentiment conditional on the numerical rating.

\begin{figure}
	\includegraphics[width=.75\textwidth]{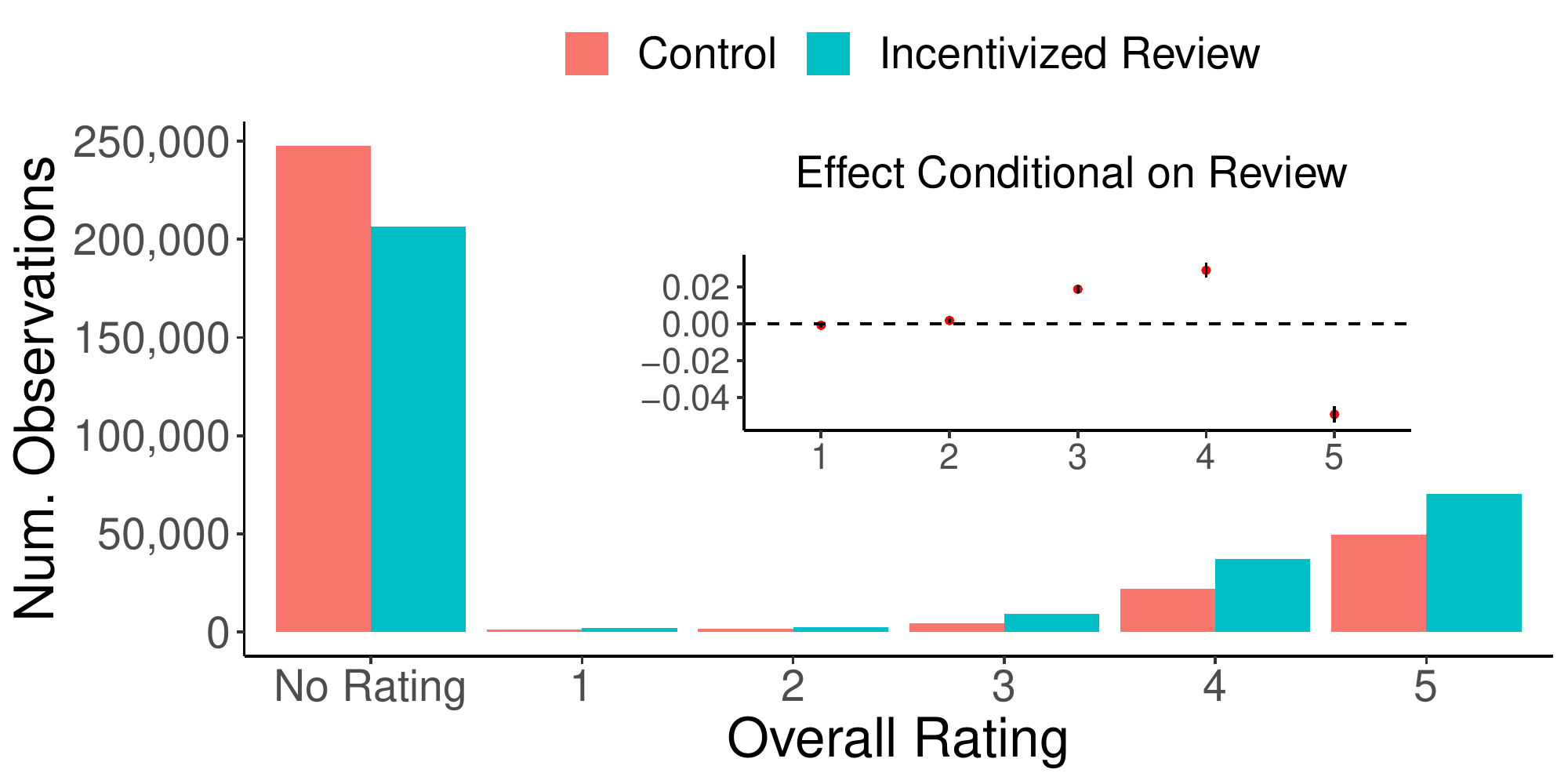}
	\centering
	\caption{Distribution of Ratings for Focal Stay}
	\label{fig:ratings_dist_comparison}
	\footnotesize \emph{Notes:} \raggedright  Comparison of the distribution of ratings left in the treatment group and the control group during the experiment. We only include the first review left for each listing. The inset plot contains the treatment effect and 95\% confidence interval conditional on a rating being submitted.
\end{figure}

Figure \ref{fig:ratings_dist_comparison} compares the distributions of numerical ratings left by guests staying at treated listings, and guests staying at control listings. The first thing that is apparent from the figure is that the treatment is effective at increasing the rate at which guests leave reviews: the treatment increases the review rate by \reviewrateeffect{} percentage points, from \reviewratectr{}\% to \reviewratetrt{}\% ($p < 2.2 \times 10^{-16}$). Because of this increase in review rate, before conditioning on a review being left, the treatment also increased the number of 5-star reviews (\reviewratefive{} pp; $p < 2.2 \times 10^{-16}$), 4-star reviews (\reviewratefour{} pp; $p < 2.2 \times 10^{-16}$), 3-star reviews (\reviewratethree{}  pp; $p < 2.2 \times 10^{-16}$), 2-star reviews (\reviewratetwo{} pp; $p < 2.2 \times 10^{-16}$), and 1-star reviews (\reviewrateone{} pp; $p < 2.2 \times 10^{-16}$). 

The majority of reviews have a five star rating, which is a fact that has previously been documented in \citep{fradkin2020reciprocity}. The high rate of five star reviews has been though to in part reflect bias in the reputation system of Airbnb. We next measure whether our intervention reduces this bias by changing the distribution of ratings.

The inset in Figure \ref{fig:ratings_dist_comparison} shows that conditional on a review, ratings of treated listings were lower than ratings of control listings; the treatment caused the average rating left by guests to drop by \condavgrate{} stars, from \condavgctr{} to \condavgtrt{} ($p < 2.2 \times 10^{-16}$). The treatment had a lower rate of five star reviews and a higher rate of 2 - 4 star reviews. In other words, while the treatment led to an across-the-board increase in the number of reviews at all levels, the increase was larger for lower ratings than for higher ratings. 

We also measure the effects of the treatment on the sentiment of review text. We describe our methodology for text classification and the details of our results about sentiment in \autoref{app:text}. Reviews for treated listings are more negative than for control listings. In particular, \meanposctr{}\% of reviews in the control group and \meanpostrt{}\% of reviews in the treatment are classified as positive ($p < 3.9 \times 10^{-9}$). These differences in text sentiment disappear once we condition on the star rating, which suggests that the effects on star ratings and review text are consistent with each other.  

Next, we consider the characteristics of reviewed transactions across treatment and control groups. These characteristics are important since they reveal what types of experiences incentivized reviews reflect. We conduct two comparisons. First, we compare the characteristics of reviewed transactions across treatment and control groups. This tells us whether reviewed experiences are meaningfully different across groups. Second, we compare the characteristics of reviewed and non-reviewed transactions in the treatment group. This reveals which experiences are not captured by reviews in the treatment.

\autoref{fig:typesofrevs} displays the differences in characteristics between reviewed trips in the treatment and control groups. Reviews in the treatment group tend to be lower value, whether measured by nights, bedrooms, or prices. At the same time, the customer complaint rate does not differ across groups, suggesting that the quality of reviewed transactions is not too different between the groups. \autoref{fig:typesofrevsnoncomp} displays the differences in characteristics between reviewed and non-reviewed trips in the treatment group. Reviewed trips are less likely to have customer complaints and have lower transaction values and prices per night. This suggests that many lower quality transactions (expensive trips and those with customer complaints) are not reviewed even in the treatment.

Our interpretation of the above results is that the incentive induces relatively more of those with lower value and mediocre experiences to review, when they otherwise would not have. This addition of these reviews should reduce the selection bias found in reviews for treated listings. However, many low quality transactions are not reviewed, even in the treatment. In the next section, we study how induced reviews affect market outcomes.

\section{Effects of Incentivized Reviews on Market Outcomes}\label{sec:effectonoutcomes}

The platform's objective in incentivizing reviews is to improve market outcomes. In this section, we measure these effects and relate them to our theoretical framework. We begin by showing that reviews have transitory effects on the number of transactions on the platform and that they do not affect overall transaction value or nights booked. We then show that transaction quality is not improved and that, if anything, it falls. Finally, we show how market structure and the design of the reputation system help to explain our findings.

One path that we avoid taking in our main analysis of the treatment's effects is conditioning on whether a review happens or on the rating of the review. Although this approach has intuitive appeal, it suffers from severe omitted variable bias. Whether a listing is reviewed and the rating of the review may be correlated with many other characteristics of the listing which are difficult to control for. These include photos and text, the communication style of the host, the availability of the listing, the quality of competition, and the balance of supply and demand.

\subsection{Effects on Demand for a Listing}
We begin by measuring the effects of being assigned to the treatment on quantities post-transaction. Our main empirical specifications are linear regressions of the following form:
\begin{equation}\label{eqn:main}
	y_{l, h} = \beta_{0} + \beta_{1} T_{l} + \epsilon_{l}
\end{equation}
where $T_{l}$ is an indicator for whether the listing, l, had a guest who was sent a treatment email offering a coupon in exchange for the review and $y_{l, h}$ is a listing outcome such as the number of transactions at a time horizon, h. The time horizon encompasses the time between the checkout for the focal transaction in the experiment and h days afterward. 

\begin{figure}
	\includegraphics[width=1\textwidth]{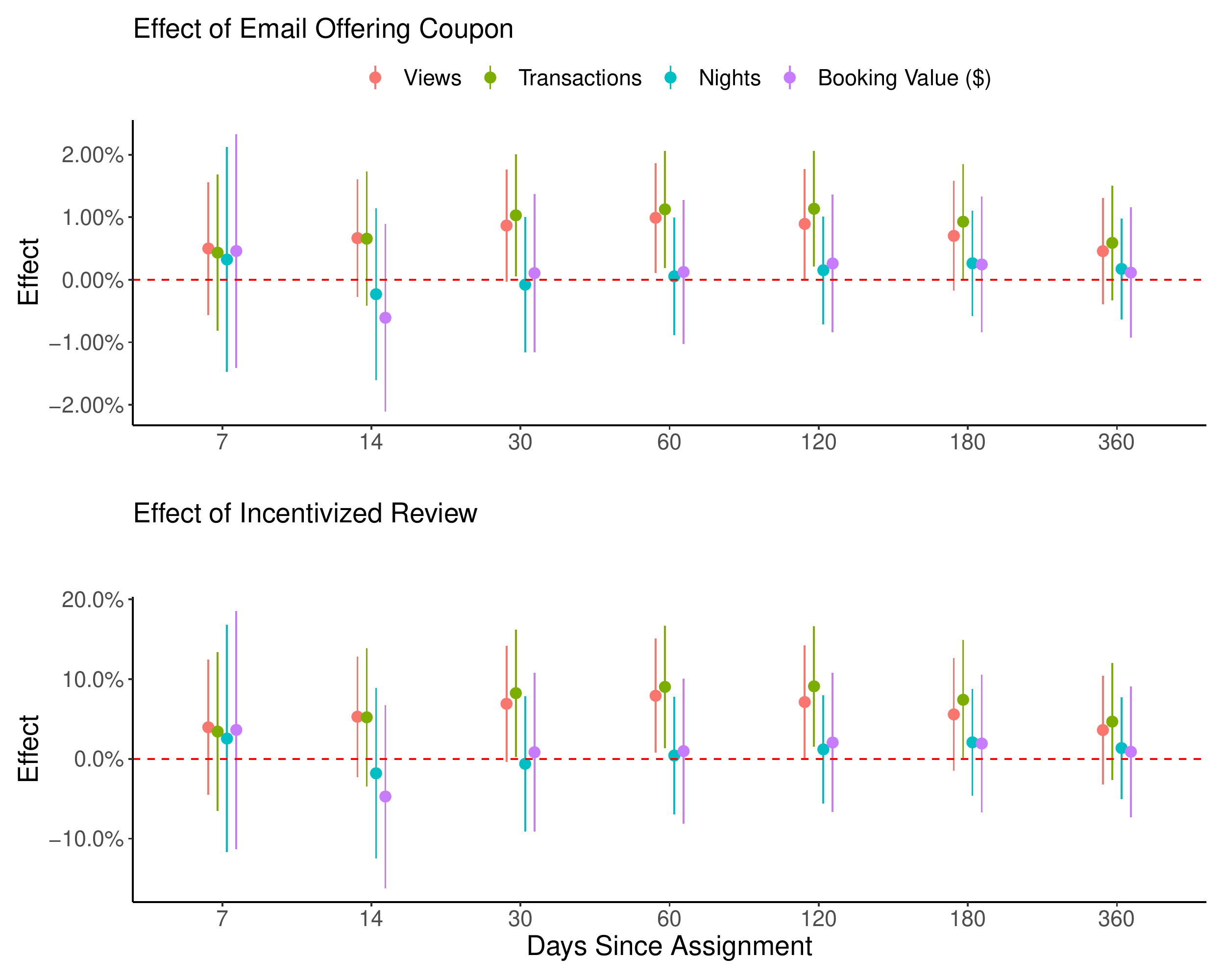}
	\centering
	\caption{Cumulative Effects of Treatment on Listing Outcomes}
	\label{fig:effectsovertime}
	\footnotesize \emph{Notes:} \raggedright The figure plots the effects and 95\% confidence intervals from \autoref{eqn:main}, where coefficients are transformed into percent terms by dividing by the intercept. Each point represents the effect of a listing's guest receiving a treatment email on an outcome measured cumulatively between the checkout for the focal transaction and days since assignment. Standard errors are calculated using robust standard errors and the delta method for the ratio of the treatment coefficient and intercept.  
\end{figure}

We consider four listing level outcomes, views after the focal checkout, transactions initiated after the focal checkout, nights for transactions initiated after the focal checkout, and the price to guests of transactions initiated after the focal checkout. \autoref{fig:effectsovertime} displays the results in percent terms. Turning first to views, we see that treated listings receive up to 1\% more views, with the effect peaking between around 60 days and then diminishing. Similarly, we see that transactions also increase by about 1\% after assignment, with the effect peaking at 120 days. On the other hand, the total nights of stay and booking value exhibit effects close to 0, which are statistically indistinguishable from 0. The effects in percentage terms shrink as the horizon expands, which reflects the temporary effects of the treatment. In \autoref{app:additionalresults} we find that the effect on reservations comes from the intensive margin and that the estimates remain similar when adding controls. 

We use a two-stage least squares estimator to translate the effect of a guest receiving an email into the local average effect of an incentivized review. To do so, we must make the standard assumptions for experiments with non-compliance. First, that the coupon email does not change the reviewing behavior of those who would have reviewed regardless of the email (the always takers). Second, that the email did not dissuade anyone from reviewing (no defiers). 
We estimate the following equation:
\begin{equation}\label{eqn:mainiv}
	y_{l, h} = \beta_{0} + \beta_{1} R_{l} + \epsilon_{l}
\end{equation}
where $R_{l}$ takes the value of 1 if the listing, l, was reviewed for the focal transaction in the experiment and where the instrument is the treatment assignment in the incentivized review experiment. Note that in the case with no covariates, the 2SLS estimate of a review will simply scale the estimate in \autoref{eqn:main} by one over  \reviewrateeffect{} percentage points, the causal effect of the coupon email on the review rate.

The second panel of \autoref{fig:effectsovertime} displays the two-stage least squares estimates measuring the effects of an incentivized review. We find that the reviews generate more attention and transactions for listings. Specifically, the effect at 120 days after the focal checkout is \effecatelateview{}\% on views and \effecatelatetrans{}\% on transactions, which represents an additional \effecatelatetranslevel{} transactions. Furthermore, the fact that views and transactions increase by similar percentages, suggests that the effect of a review comes from increased clicks from the search page to the listing page. In \autoref{app:views} we show that this effect is driven by the fact that the number of reviews is displayed on the search page rather than by changes to the algorithmic ranking of a listing.

Even though transactions increase, the number of nights, which represents the total quantity sold, remains constant. This suggests that the presence of an incentivized review changes the \textit{types} of trips that occur. We investigate this by analyzing a transaction level dataset. In particular, we take the set of all post-assignment transactions that are booked within 120 days of assignment and estimate regressions at the trip level, with standard errors clustered at the listing level. 

Column (1) of \autoref{tab:tripchar} shows that nights per trip fall by 1\% for treated listings. Column (2) shows a similarly sized reduction in trip revenue, although the effect is not statistically significant. Columns (3) and (4) show statistically insignificant effects on price per night and lead time, which is the number of days between a reservation and checkin. 

\begin{table}
	\caption{Effects of Treatment on Trip Characteristics}
	\label{tab:tripchar}
	\centering
	\footnotesize
		
\begin{tabular}{lcccc}
\toprule
 & Nights Per Trip & Trip Revenue & Price Per Night & Lead Time (Days)\\
 & (1) & (2) & (3) & (4)\\
\midrule (Intercept) & 4.207$^{***}$ & 396.5$^{***}$ & 103.7$^{***}$ & 17.29$^{***}$\\
  & (0.0098) & (1.340) & (0.3128) & (0.0384)\\
Treatment & -0.0403$^{**}$ & -3.473 & -0.4688 & 0.0272\\
  & (0.0136) & (1.882) & (0.4403) & (0.0543)\\
  &   &   &   &  \\
R$^2$ & $7.84\times 10^{-6}$ & $7.37\times 10^{-6}$ & $4.68\times 10^{-6}$ & $4.79\times 10^{-7}$\\
Observations & 2,389,288 & 2,389,288 & 2,389,288 & 1,892,755\\
\bottomrule
\end{tabular}

		\vspace{.1in}
	\footnotesize \raggedright \emph{Notes:} This table displays regressions at a transaction level of transaction characteristics on the treatment. All transactions for listings in the experiment that occur within 120 days of the checkout of the focal stay are considered. The regression for lead time includes fewer observations since we considered only trips for which the checkin occurred within 120 days. Standard errors are clustered at the listing level.
\end{table}

To summarize, the net effect of reviews on listing quantity and revenue is close to 0. This is true even though treated listings get more views and transactions. One reason that more transactions do not necessarily translate into more nights is that the capacity of Airbnb listings is limited. In particular, unlike in goods markets, only one buyer can book a listing per night. As a result, increased views can only increase quantity if they result in bookings for marginal nights. Another response by sellers could be to increase nightly prices due to being reviewed but they do not. As shown in \citet{huang2021seller}, sellers on Airbnb are often inattentive or constrained in changing prices when responding to demand fluctuations.

\subsection{Effects on Transaction Quality}
Even if incentivized reviews have small effects on average demand, they may affect the quality of matches that occur. This could happen if, for example, the review text contained information that helped potential guests pick listings. To test for this, we construct transaction level customer satisfaction proxies for transactions post-treatment. More concretely, for each listing, $l$, we consider all transactions that occur within 120 days after the checkout of the focal stay, do not have a cancelation, and have an observed payment. For this sample of observations, we measure customer complaints, reviews, and customer return rates. Customer return rates are measured by the number of subsequent nights on the platform for guests staying at the listing post-treatment.\footnotemark{}


\footnotetext{User return rates to the platform have been used as a measure of customer satisfaction in \citet{nosko2015limits} and \citet{farronato2020consumer}.}

\autoref{tab:matchq} displays the results of the transaction quality regressions. In particular, the service complaint rate (Column 1) is not statistically different for post-treatment transactions between treated and control listings. The review rate increases, but this is caused in part by the fact that guests of treated listings are eligible for the coupon until the listing has a first review. In particular, for treated listings who do not get a review in the focal transactions, the next guest could also receive an incentivized review offer if they did not review within 7 to 10 days. Column (3) shows that conditional on a subsequent review, the rating is worse in the treated group. This is consistent with the fact that some treated listings may receive an incentivized review from subsequent transactions, rather than the first one and also with a worse match quality in the treatment group.\footnotemark{} The addition of covariates does not substantively affect these results (\autoref{tab:matchqctr}).

\footnotetext{Another reason, proposed by \citet{hui2021and}, is that low ratings may be autocorrelated due to belief updating dynamics that affect review rates.}

Column (4) of \autoref{tab:matchq} displays the effects on the return propensities of guests who stay at listings after the focal transaction. Guests to treated listings stay for fewer nights after a transaction than guests to control listings. The effect is statistically significant and represents a 1.6\% decrease at nights. 

This effect on guest's subsequent platform usage can be due to one of two mechanisms. The first is that incentivized reviews cause worse matches and cause guests to use the platform less as a result. The second is that induced reviews induce matches with different types of guests, who are also less likely to use the platform afterwards. To investigate this further, in column (5), we add controls for guest and trip characteristics. If different types of guests are induced to stay due to incentivized reviews, then these controls could capture these differences. We still detect a statistically significant effect of the treatment, although the point estimate is smaller. 

The small negative effect in column (5) of \autoref{tab:matchq}  is consistent with a theoretical possibility that incentivized reviews actually cause \textit{worse} matches. In particular, as described in \autoref{sec:theory}, incentivized five star reviews may be induced for low-quality listings. As a result, low-quality listings may appear to subsequent guests as higher quality and may thus cause worse matches. We conclude that, if anything, users who stay at treated listings may have a worse experience.

\begin{table}
	\caption{Effects of Treatment on Transaction Quality}
	\label{tab:matchq}
	\footnotesize
	\centering
		
\begin{tabular}{lccccc}
\toprule
 & Complaint & Reviewed & Star Rating & \multicolumn{2}{c}{Guest Nights}\\
 & (1) & (2) & (3) & (4) & (5)\\
\midrule Constant & 0.0101$^{***}$ & 0.6475$^{***}$ & 4.529$^{***}$ & 5.591$^{***}$ &   \\
  & (0.0001) & (0.0006) & (0.0014) & (0.0217) &   \\
Treatment & $-6.52\times 10^{-5}$ & 0.0048$^{***}$ & -0.0060$^{**}$ & -0.0766$^{**}$ & -0.0548$^{*}$\\
  & (0.0001) & (0.0009) & (0.0020) & (0.0296) & (0.0245)\\
  &   &   &   &   &  \\
R$^2$ & $1.06\times 10^{-7}$ & $2.52\times 10^{-5}$ & $1.53\times 10^{-5}$ & $6.48\times 10^{-6}$ & 0.20805\\
Observations & 2,431,085 & 2,431,085 & 1,579,132 & 2,431,085 & 2,431,085\\
  &   &   &   &   &  \\
Controls & No & No & No & No & Yes\\
Guest Region FE &  &  &  &  & $\checkmark$\\
Checkout Week FE &  &  &  &  & $\checkmark$\\
Num. Nights FE &  &  &  &  & $\checkmark$\\
Num. Guests FE &  &  &  &  & $\checkmark$\\
\bottomrule
\end{tabular}

		\vspace{.1in}

		\footnotesize \emph{Notes:} \raggedright  This table displays regressions measuring the effects of the treatment (the guest receiving an email with an offer of a coupon in exchange for a review) on measures of transaction quality. The set of transactions considered for this regression includes all transactions for which the checkout date was between the checkout date of the focal transaction and 360 days after. `Complaint' refers to whether a guest submitted a customer service complaint to Airbnb, `Reviewed' refers to whether the guest submitted a review, `Star Rating' refers to the star rating of any submitted reviews, `Guest Reservations' and `Guest Nights' refer, respectively, to the number of transactions and nights of a guest in the 360 days post checkout. Control variables in (5) include the log of transaction amount, the number of times the guest has reviewed and reviewed with a five star ratings in the past, the prior nights of the guest, whether the guest has an about description, and guest age on the platform.
\end{table}

The treatment may also have affected the behavior of hosts. We measure whether hosts change their listing page in response to review related information. Specifically, we measure whether the number of photos or the length of a listing's description changed due to the treatment. \autoref{tab:moralhazard} shows precisely estimated null effects, meaning that, at least in terms of how hosts advertise their listing, there is no effect. 

\subsection{Why Do Incentivized Reviews Have Small Effects?}\label{sec:whysmalleffect}

In this section we document two reasons why incentivized reviews have have small effects on demand. The first is that the listings in our sample are able to generate transactions even without a review. As a result, their other transactions can generate a first review, even if the focal transaction in the experiment did not. These first reviews typically arrive quickly and have a similar ratings distribution to other reviews. The second reason is that star ratings (as opposed to review text) are only displayed after a listing has three reviews, by which point differences between incentivized and control reviews average out. 


Listings in our sample have been able to get at least one booking without a review. This means that, at least for some guests, the presence of a first review is not pivotal in their choice. One reason this occurs is that there are supply-constrained markets with few other options for guests.\footnotemark{} As a result, guests are shown listings without reviews and sometimes book these listings.

\footnotetext{We can measure the degree of supply-constraints using the ratio of the number of inquiries divided by the number of listings contacted by market during the time that the experiment was conducted. The average listing in our experiment is booked in a market where the tightness (\tightnessexp{}) is much higher than the tightness in a typical market (\tightnessall{}).}

\sharepreexp{}\% of listings in the experiment have more than one booking prior to the checkout of the focal trip. Each of these additional bookings offers an opportunity for the listing to receive a review and these opportunities add up. We find that \ratefirsteverctr{}\% of listings in the control group eventually receive a review, while \ratefirstevertreat{}\% do so in the treatment group. This \ratefirsteverdiff{}\% difference is less than half as large as the effect on the treatment for the focal transaction (13\%). 

\begin{figure}
	\includegraphics[width=.75\textwidth]{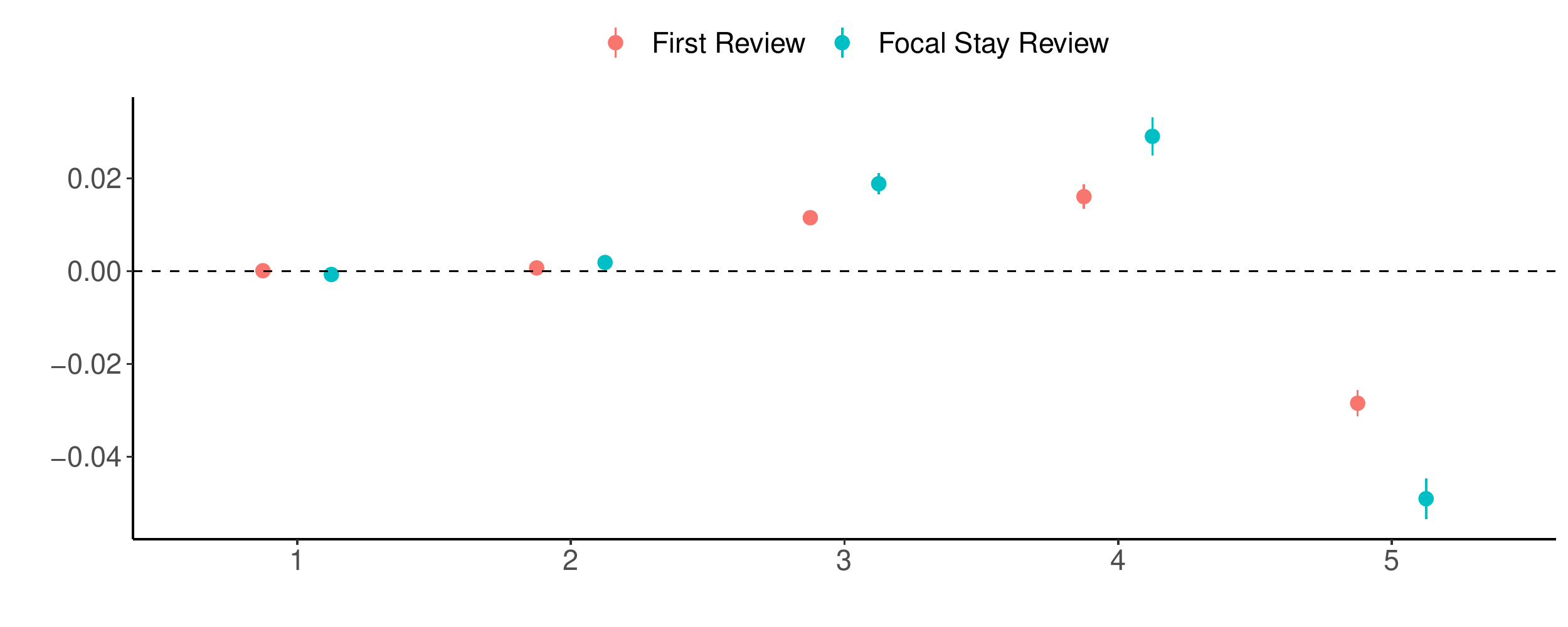}
	\centering
	\caption{Effect on Review Ratings (Conditional on Review)}
	\label{fig:reviews_dist_cond_effect}
	\footnotesize \emph{Notes:} \raggedright  The figure plots the estimate and 95\% confidence interval for differences in the share of reviews with each rating type. `Focal Stay Review' refers to any review that occurred for the first transaction for a listing that was eligible for the experimental treatment. `First Review' refers to the first review ever received by a listing.  
\end{figure}

The ratings for first reviews occurring outside of the focal stay are more similar between the treatment and control groups than those that occur for the focal transaction. \autoref{fig:reviews_dist_cond_effect} plots the differences in ratings between treatment and control group for reviews coming from the focal transaction and for any first reviews. The effect sizes for first reviews are smaller in magnitude for each rating. This smaller difference in ratings is likely to contribute to the small effects on demand and matching that we find.

\begin{figure}
	\includegraphics[width=.75\textwidth]{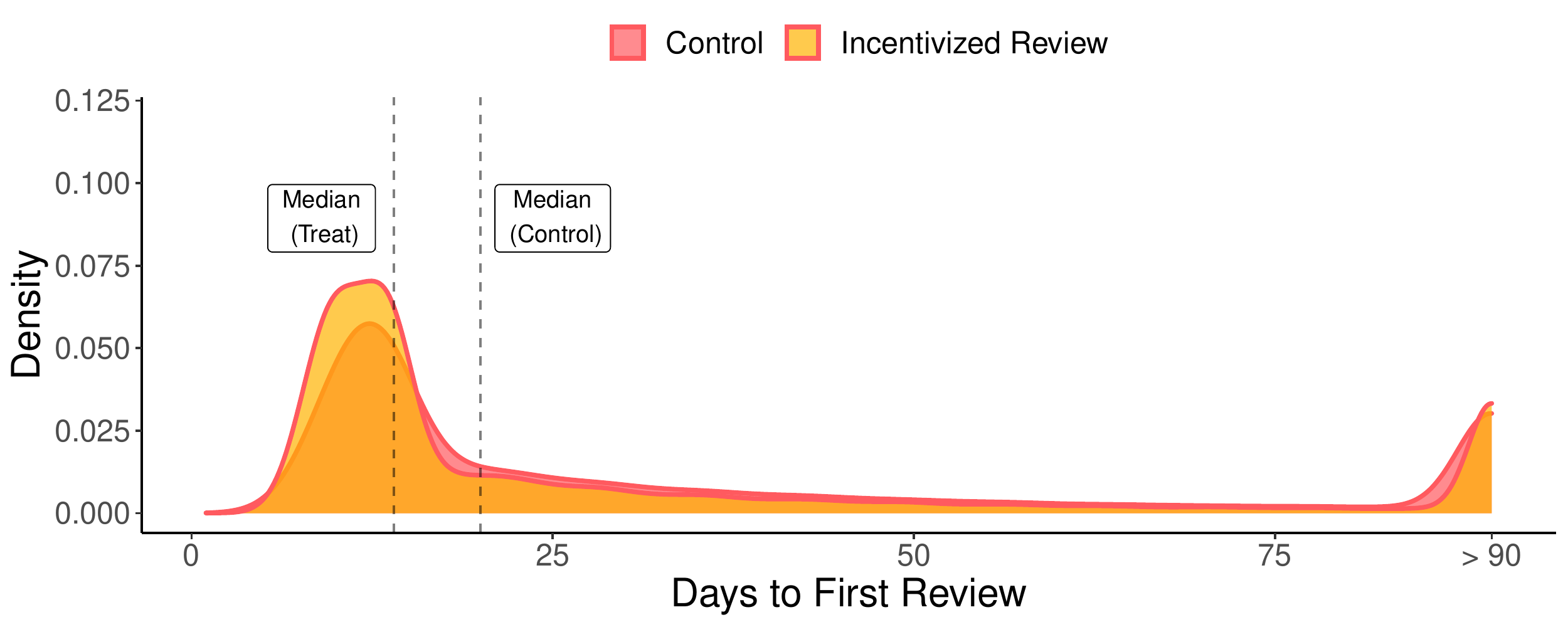}
	\centering
	\caption{Distribution of Days to First Review}
	\label{fig:reviews_time_1st}
	\footnotesize \emph{Notes:} \raggedright  The figure plots the distribution of the time of arrival for the first review for treated and control listings. The time is calculated as the difference in days between the date of the arrival of the first review and the checkout of the transaction for which the experimental assignment occurred. 
\end{figure}


Another reason why the effects of incentivized reviews are muted is the manner in which Airbnb displays reviews. Ratings are not shown for every review but are instead averaged and rounded to the nearest half star. Rounding to half a star is a common design online and is used by Yelp, Etsy, and Facebook Marketplace. Furthermore, while review text is displayed even after the first review, average ratings are only shown after 3 reviews. This means that the differences in ratings between the treatment and control groups are attenuated through averaging. 

\begin{figure}
	\includegraphics[width=.75\textwidth]{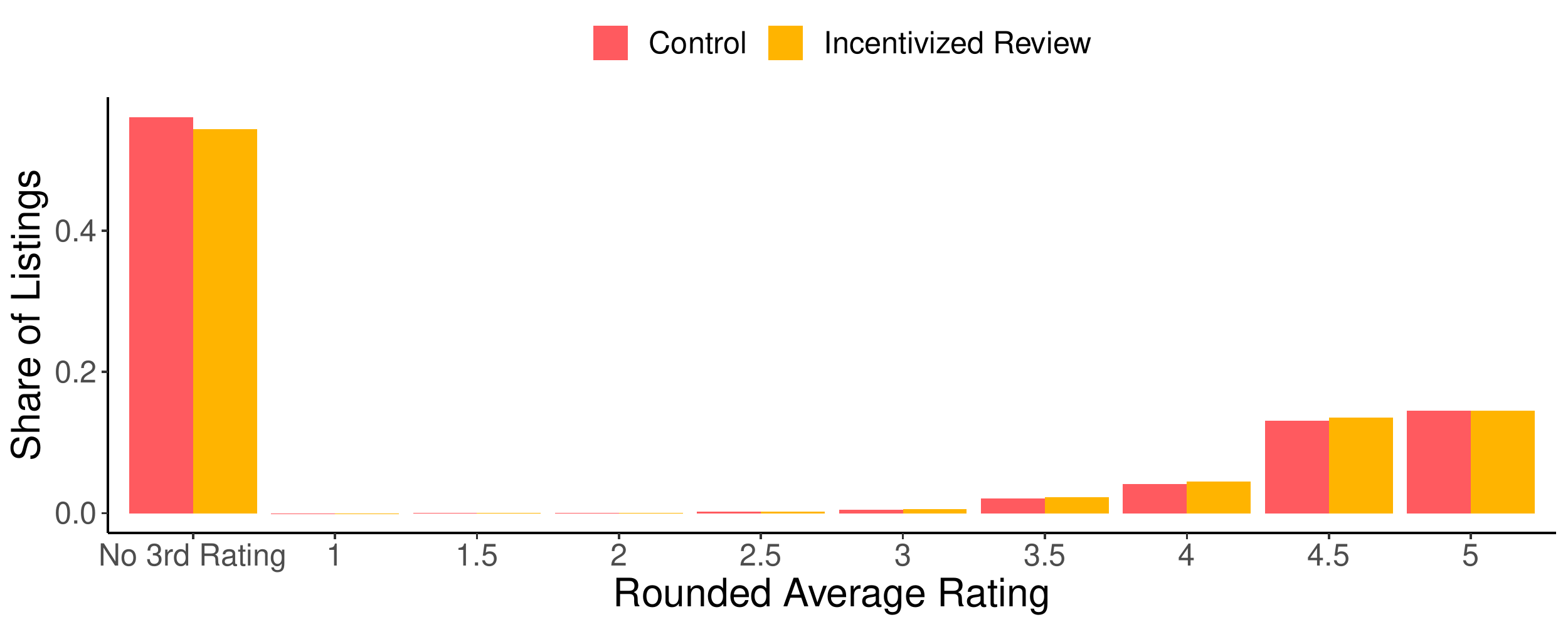}
	\centering
	\caption{Rounded Average Rating After Three Ratings}
	\label{fig:reviews_dist_3rd}
	\footnotesize \emph{Notes:} \raggedright  The figure plots the distribution of the rounded average of the first three ratings from any transaction in the treatment and control groups. 
\end{figure}

\autoref{fig:reviews_dist_3rd} shows that treatment listings are \ratethirdeverdiff{} percentage points more likely to get to three ratings, which is a much smaller difference than for the first reviews. The average rounded ratings also exhibit much smaller differences in shares than the first ratings.   

To summarize, even control listings typically obtain first reviews. These first reviews come quickly after experimental assignment and exhibit broadly similar ratings. As a result, first incentivized reviews would need to have effects in a relatively short amount of time for them to substantially affect demand and matching outcomes. Furthermore, since ratings, as opposed to text, are only shown after three reviews incentivized ratings are averaged out in aggregate.

\subsection{Large Heterogeneous Treatment Effects Do Not Explain the Small Average Effect}\label{sec:het}

Another potential explanation for the small average treatment effects is that the incentivized reviews have highly heterogeneous effects. Some listings, such as those on the margin of getting additional bookings, may benefit a lot from an incentivized review while others who would have gotten reviewed regardless may primarily face downside risk. Another hypothesis is that five star reviews benefit listings while lower star reviews hurt listings in a way so that their effects average out. We now investigate both of these heterogeneities and fail to find evidence that large heterogeneous effects drive our main results. 

In order to test for heterogeneity with regards to benefits from a review, we need a variable that proxies for the benefit to a listing of a review. One candidate for such a variable is the predicted future demand for a listing. We'd expect that a review benefits listings who would have otherwise done poorly on the platform and may not benefit or even hurt listings who are predicted to do well. We construct this proxy in three steps. 

First, we select a similar but auxiliary sample on which to train the prediction model. This avoids having to conduct sample splitting procedures as in \citet{guo2021machine}, who propose a similar way to reduce variance and estimate heterogeneous treatment effects for the purpose of analyzing digital experiments. Our sample consists of previously non-reviewed listings who were reviewed within 9 days of the checkout, and were thus not eligible for our experiment. Intuitively, this is a similar population of listings and so the covariates that predict success on the platform should be similar to those of the experimental sample.

Second, we use a estimate a linear regression with listing outcomes as dependent variable and pre-checkout covariates, market, and location fixed effects as control variables. This regression gives us the coefficients in a linear regression on the included variables. Third, we apply these coefficients to the experimental sample in order to create a prediction for each listing in the sample of the listing outcomes.

To test for heterogeneity, we estimate a regression of the following form (as suggested by \citet{lin2013agnostic}):\footnotemark{}
\begin{equation}\label{eqn:het}
	y_{l} = \beta_{0} + \beta_{1} T_{l} + \beta_{2} X_{l} + \beta_{3} T_{l} (X_{l} - \bar{X}) + \epsilon_{l}
\end{equation}

In the above regression, $y_{l}$ is a listing outcome (reservations, nights, and booking value) within 120 days of the focal stay checkout and $T_{l}$ is the treatment indicator, while $X_{l}$ is the prediction of the outcomes and $\bar{X}$ is it's average. The interaction coefficient, $\beta_{3}$ is our main coefficient of interest.
\footnotetext{\citet{lin2013agnostic} shows that this specification allows $\hat{\beta_{1}}$ to be consistent for the average treatment effect even in the presence of covariates.}

\autoref{tab:heterogpredicted} displays the results from \autoref{eqn:het}. We see that predicted nights is indeed a good proxy since the coefficient is higher than .5 and the $R^2$ rises from approximately 0 to between 13\% and 19\% depending on the regression. Nonetheless, the interaction term is statistically insignificant and small in magnitude. As a result, heterogeneity with regards to potential success on the platform does not explain the small average effects of the treatment.\footnotemark{}

\footnotetext{We also conduct a more standard analysis of heterogeneity in \autoref{tab:heterogsamp}, where we interact the treatment with one covariate at a time for six co-variates of interest: listing age, Superhost status, multi-listing host, host gender, price per night, and whether the host was booked more than once prior to the focal stay checkout. We only detect statistically significant heterogeneity with regards to nightly price. We find that the treatment increases reservations less for more expensive stays.}

\begin{table}[hptb]
	\caption{Heterogeneity by Predicted Outcomes}
	\label{tab:heterogpredicted}
	\begin{center}
	\footnotesize
	
\begin{tabular}{lccc}
\toprule
 & Reservations & Nights & Booking Value\\
 & (1) & (2) & (3)\\
\midrule (Intercept) & -0.3024$^{***}$ & 0.5944$^{***}$ & 91.10$^{***}$\\
  & (0.0271) & (0.0847) & (17.05)\\
Treatment & 0.0349$^{**}$ & 0.0287 & 4.260\\
  & (0.0152) & (0.0629) & (7.942)\\
Predicted Reservations & 0.5431$^{***}$ &    &   \\
  & (0.0042) &    &   \\
Treatment $\times$ Predicted Reservations (Demeaned) & 0.0053 &    &   \\
  & (0.0066) &    &   \\
Predicted Nights &    & 0.5970$^{***}$ &   \\
  &    & (0.0039) &   \\
Treatment $\times$ Predicted Nights (Demeaned) &    & 0.0093 &   \\
  &    & (0.0062) &   \\
Predicted Booking Value &    &    & 0.6630$^{***}$\\
  &    &    & (0.0082)\\
Treatment $\times$ Predicted Booking Value (Demeaned) &    &    & 0.0021\\
  &    &    & (0.0102)\\
Horizon & 120 Days & 120 Days & 120 Days\\
  &   &   &  \\
Observations & 640,936 & 640,936 & 640,936\\
R$^2$ & 0.16055 & 0.13454 & 0.18840\\
\bottomrule
\end{tabular}

	\end{center}
\footnotesize \emph{Notes:} \raggedright  This table displays the regression estimates from \autoref{eqn:het}, where the outcome is reservations, nights, and booking value within 120 days of the focal checkout. Predicted reservations, nights, and booking values are calculated using the procedure described in \autoref{sec:het}. Note that the number of observations in this regression is lower than in the others since some uncommon fixed effect values in the experimental data were not present in the training data and some covariates were missing for some of the observations.
\end{table}

Next, we investigate whether heterogeneous effects due to some listings receiving good reviews and other listings receiving bad reviews can explain our results. Note that we cannot take an approach similar to the one above, since it's difficult to predict ratings and since submitted ratings are endogenous. Instead, we turn to a calibration exercise. We know from \autoref{sec:effectsonreviews} that the treatment increased the likelihood of a review with rating, r, by an amount $z(r)$. If we also knew the causal effect of a review with rating r, $\tau(r)$ relative to no review on an outcome, $Y$, then we could calculate the average treatment effect using the following equation:

\begin{equation}\label{eqn:predicteffect}
 	E[Y|T=1] - E[Y |T = 0] = \sum_{r \in \{1, 2, 3, 4, 5\}} \tau(r) z(r)
\end{equation}

Although we don't know $\tau(r)$, we can use multiples of the observational estimate as a benchmark. In particular, suppose we use a linear regression to predict future demand as a function of the star rating, and treat the coefficient on the rating as an estimate of $\tau(r)$. \autoref{fig:corrrating} displays the observational estimates of the effect of a review in the control group on 120 day nights and revenue. We see that five star reviews are associated with much higher demand relative to no review, while one, two, and three star reviews are associated with much lower demand. Note that these estimates are likely to be biased upward in magnitude even after adding controls, since the rating is correlated with factors observable to guests but not to the econometrician. To account for this, we can also test the sensitivity of our calibration to estimates of $\tau(r)$ which are shrunken towards 0 by a factor $k<1$.

We plug in the observational estimates with controls into \autoref{eqn:predicteffect} and obtain a calibrated estimate of \diffpredictonenights{} for the treatment effect on nights. This estimate is much larger than the regression estimates of \regestimatenights{} on nights and is outside of the 95\% confidence interval. We then consider shrinkage factors of .5 and .25, for which we find predicted effects on nights of \diffpredictfnights{} and \diffpredicttfnights{} respectively, which are still larger than the estimated treatment effects.\footnotemark{}

\footnotetext{Using shrinkage factors of 1, .5, and .25, we find expected effects on revenue of \$\diffpredictonerev{}, \$\diffpredictfrev{} and \$\diffpredicttfrev{} respectively. The point estimate of the treatment effect is, in contrast, \$\regestimaterev{}, although it is less precisely estimated.}

Both exercises in this subsection have failed to find that heterogeneity in effects can explain the small and statistically insignificant average treatment effects of the treatment on nights and revenue. As a result, we conclude that the effects of incentivized first reviews on listing demand are typically small and that naive observational estimates of the effects of reviews are mostly explained by selection bias. 

\section{Discussion}\label{sec:conclusion}

We studied when and whether reviews are underprovided, by analyzing a large-scale experiment in which buyers were incentivized to submit additional reviews. We found that the incentive was successful in inducing reviews. These incentivized reviews exhibited lower ratings, consistent with the presence of selection bias in the reviewing process. However, although the treated group was reviewed faster with reviews that had lower ratings, there were small effects on demand, revenue, and match quality. 

We documented two main reasons for these small effects. First, listings were able to obtain reviews quickly even if they were not reviewed for the focal transaction for which the incentive was offered. Second, since ratings were only displayed as rounded averages after a listing had at least three reviews, any differences between treatment and control reviews were attenuated.  

Our findings have implications for the understanding of the role of reviews in a marketplace and for the design of reputation systems. With regards to the role of reviews, we show that for the setting of Airbnb, faster first reviews do not have large effects. We argue that this is due to the structure of the market and reputation system. If listings are able to obtain transaction without reviews, then they have multiple opportunities to get reviews and to accumulate reputation. As a result, the effect of a marginal review is likely to be small. This small effect of reviews is further exacerbated by the fact that ratings were displayed as rounded averages and only after at least three reviews were present.  

Our results are specific to Airbnb's reputation system design. In this design, ratings are only shown after three reviews and this may reduce the dependence of a listing's success on the marketplace on the first review. A reduction in path dependence could be good if marketplaces are worried about false negative reviews, but may be bad if there are many false positive reviews on the marketplaces. The optimal rounding of ratings is also on open question in reputation system design. Subsequent to our experiment, Airbnb changed it's rounding to two decimal places from half a point (\citet{ratingrounding}). This change may be complementary to incentivized reviews since it makes it easier for guests to distinguish between the ratings of different listings.

There could also be alternative incentivized review programs which are more suited toward improving market outcomes. Our treatment induced reviews for a specific set of transactions. Other incentivized review policies could be better targeted and have larger payouts. For example, the platform could target incentives towards sellers with uncertain quality and it could make these incentives have higher stakes.  The accumulated set of incentivized reviews will likely have a much larger effect on demand for that seller then the single incentivized review produced with our experiment. Alternatively, the incentive could be conditioned on the review text being informative. These more informative reviews may have different effects from the induced reviews studied in our work. 

Lastly, the incentivized review policy that we study is not well suited toward solving the cold-start problem. In particular, the policy targets the set of listings who did not need a review to obtain their first transaction. In order to solve the cold-start problem, the platform would need alternative market designs. These include subsidizing transactions to sellers without prior transactions, boosting new sellers in search, and hiring `mystery shoppers' to examine the quality of new inventory. Whether the policies would be successful is an open question that we leave for future work. 

\clearpage
\bibliography{reviews_incent}
\newpage
\onehalfspace


\appendix

\counterwithin{figure}{section}
\counterwithin{table}{section}

\clearpage
\section{Theoretical Model}\label{app:theory}
Whether the platform should incentivize reviews depends on whether these reviews improve outcomes on the platform. In this section, we describe a theoretical framework which clarifies the conditions under which incentivize reviews increase demand and the utility of buyers. The framework is a simplified version of \citet{acemoglu2019learning}, which characterizes the speed of learning in review systems and shows that review systems with a higher speed of learning increase expected the expected utility of buyers. In this theoretical framework, the degree to which incentivized reviews improve buyer utilities is a function of the informativeness of the review system, which is a measure of the extent to which buyer beliefs about quality correspond to the true quality of a listing. This informativeness is a function of both the extent to which ratings correlate with quality and the extent to which buyer's beliefs about ratings correspond to rational expectations.

Suppose that a buyer is randomly matched with a seller. The seller has a true underlying quality $Q \in \{0, 1\}$ and an associated review outcome $r \in \{-1, 0, 1\}$, where -1 corresponds to a negative review, 0 to no review, and 1 to a positive review. The utility of buyer, i, for listing, $l$, is:
\begin{equation}
	u_{il} = \theta_{i} + Q_{l} - p
\end{equation}
In the above equation $\theta \sim F$ is the ex-ante preference of the buyer for the inside option and p is the price of the listing, which we assume to be constant. The buyer does not know the true value of $Q_{l}$ and must therefore form a guess based on the review (or lack thereof) and prior beliefs. 

The platform has a review system, $\bm{\Omega}$, which maps the history of transactions to reviews. Examples of $\bm{\Omega}$ include a review system without incentivized reviews and a review system with incentivized reviews. Let $\Omega_{l}$ be a realized outcome of the review system for listing $l$, where prior buyers have the opportunity to submit reviews. The buyer observes $\Omega_{l}$ and forms a belief $q_{i}(\Omega_{l})$ about the probability that listing $q$ has quality equal to 1. The buyer then makes a utility maximizing purchase decision:
\begin{equation}
	b_{il} = \text{arg} \max_{b \in \{0, 1\}} \textbf{1}\{b = 1\} (E_{Q}[\theta_{i} + Q_{l} - p | \Omega_{l}]) = \text{arg} \max_{b \in \{0, 1\}} \textbf{1}\{b = 1\} (\theta_{i} + q_{i}(\Omega_{l}) - p)
\end{equation}

\citet{acemoglu2019learning} show that in setups similar to this, if consumers have rational expectations and play a pure-strategy Bayesian equilibrium, the beliefs of a sequence of arriving buyers converge to the true seller quality.\footnotemark{} One boundary condition of this model is worth highlighting. If the upper bound on $\theta$ is insufficiently high, then high quality listings may get unlucky. If, for example, $\bar{\theta} < p - q(-1)$, then negatively reviewed listings will never be booked again. This will be the case even if some of those listings are of high quality and were negatively reviewed just by chance. Consequently, this model can allow for results similar to \citet{park2021fateful}, where first negative reviews have large negative effects. 

\footnotetext{\citet{acemoglu2019learning} also place restrictions on the reviewing behavior of buyers.}

Buyers' expected utilities (across preferences, quality, and realizations of the review system) can be expressed as follows, where we also assume that $\theta \in [p, 1]$ so that people prefer to purchase high quality listings but not low quality listings, $\mu$ is the share of listings that are of high quality, and that the belief function, $q_{i}$ is constant across buyers:\footnotemark{} 
\begin{equation}\label{eqn:expu}
	\begin{split}
	E_{\theta, Q, \Omega} = & \mu ( 1 - p + E_{\theta}[\theta]) \\
	+ & (1 - \mu) E_{\theta} [ - (p - \theta) P_{\Omega} [q \geq p - \theta | Q = 0] ] \\
	+ & \mu E_{\theta} [ -(1 - p + \theta) P_{\Omega} [q \leq p - \theta | Q = 1]]
	\end{split}
\end{equation}
\footnotetext{See the proof of Proposition 6 in \citet{acemoglu2019learning} for a more general formulation of this result.}

The above equation contains the key ingredients necessary for understanding the effects of a change in the reputation system. Line 1 is the utility if everyone only purchased from high quality listings. Line 2 is the false positive utility, which represents the utility loss from purchasing from a low quality listing. Line 3 is the utility lost due to false negatives, which occur when buyers do not purchase from a high quality listing. 

We now consider the effects of incentivized reviews on demand and utility in this framework. Suppose that $\bm{\Omega}_{c}$ is the control review system and $\bm{\Omega}_{t}$ is the treatment review system, and further suppose that for any stay, $\bm{\Omega}_{t}$ weakly increases review rates, but results in the same rating conditional on a review. This rules out situations where, for example, the coupon offer changes the degree of reciprocity felt by the guest. We also assume that $q(-1) < q(0) < q(1)$, meaning that positive reviews are better than no reviews and that no reviews are better than negative reviews.

Then the change in demand due to a shift from $\bm{\Omega}_{c}$ to $\bm{\Omega}_{t}$ is: 
\begin{equation}\label{eqn:demand}
	\begin{split}
		 & (\tau_{H, 1} + \tau_{L, 1}) Pr( p - q(0) > \theta > p - q(1)) \medspace - \\
		 & (\tau_{H, -1} + \tau_{L, -1}) Pr( p - q(-1) > \theta > p - q(0)) 
	\end{split}
\end{equation}

\autoref{eqn:demand} contains two lines. The first line is the increase in demand due to some listings having a positive review in the treatment, where $\bm{\Omega}_{c} (s) = 0$ and $\bm{\Omega}_{t}(s) = 1$. The mass of these listings is $\tau_{H, 1} + \tau_{L, 1}$, where H and L represent high and low quality listings respectively. This sum is identified in our experiment. For example, the number of five-star reviews increases by \reviewratefive{} pp. This sum is multiplied by the change in demand due to a positive review, which is the share of guests that would purchase if the review was high but would not purchase if there were no review. The second line of \autoref{eqn:demand} is analogous but measures the decrease in demand for listings that would have had no review in the control review system but were negatively reviewed in the treatment system. Our analysis in \autoref{sec:effectsonreviews} shows that the there is a much bigger increase in positive reviews than in negative reviews. However, it is possible that the change in demand due to the positive reviews is small, in which case the effect of incentivized reviews may be small (or negative).

The effects of incentivized reviews on expected utility are a bit more subtle. It could be the case that incentives induce the wrong types of reviews, leading to worse matches. The change in expected utility from incentivized reviews can be expressed as follows:
\begin{equation}\label{eqn:irutility}
	\begin{split}
		 & \tau_{H, 1}  E[ (1 - p + \theta) Pr( p - q(0) > \theta > p - q(1)) \medspace + \\
		 & \tau_{L, 1}  E[ (- p + \theta) Pr(p - q(0) > \theta > p - q(1)) \medspace +\\
		 & \tau_{H, -1} E[ - (1 - p + \theta) Pr( p - q(-1) > \theta > p - q(0)) \medspace +\\
		 & \tau_{L, -1} E[ - (p - \theta) Pr( p - q(-1) > \theta > p - q(0))] 
	\end{split}
\end{equation}

\autoref{eqn:irutility} contains four terms, corresponding to cases when high and low quality listings are reviewed either positively or negatively due to the treatment. The best case scenario for an incentivized review system is when the second and third lines are equal to 0, meaning that incentivized reviews increase positive review only for high quality listings and increase negative reviews only for low quality listings. But it may also be the case that incentivized reviews induce positive reviews for low quality listings. This may occur if, for example, guests value the coupon but do not want to say something negative about their stay in a review. In that case, the second line the equation would become relevant.\footnotemark{} Finally, it may be the case that a high quality listing is unlucky and gets negatively reviewed due to the treatment, a mechanism hinted at in \citet{park2021fateful}. That would correspond to line 3. The relative importance of these terms is an empirical question, which we explore in the next section.

\footnotetext{A similar mechanism is documented in \citet{muchnik2013social}, who show that randomly assigned up-votes on Reddit had large positive effects on subsequent scores.}

\clearpage

\section{Additional Results}
\subsection{Description of Review Email Dispatch During the Experiment}\label{app:daystoemail}

The intended number of days between the checkout and the emails in the experiment was intended to be 9 days for most of the sample. After March 29 of 2016, the number of days within which a review must have been submitted was changed to 7. 

In practice, the number of days varied for several reasons. First, since transactions happen around the world, the measurement of the date of the checkout and email depends on the time zone in which a checkout occurs. The email system does not perfectly take these time-zones into account. Second, at least during the period we study, stays that had partial cancelations were not fully accounted for by the email dispatch system. As an example, let's say a stay was initially booked for ten days but the guest checked out five days early. The email dispatch system still used the initial ten day booking as the basis for calculating the date of the required email. Third, the exact time of the email varied over time and across days of the week. Lastly, there seemed to be several outages of the email system during which emails were sent with a delay.

\begin{figure}
	\includegraphics[width=\textwidth]{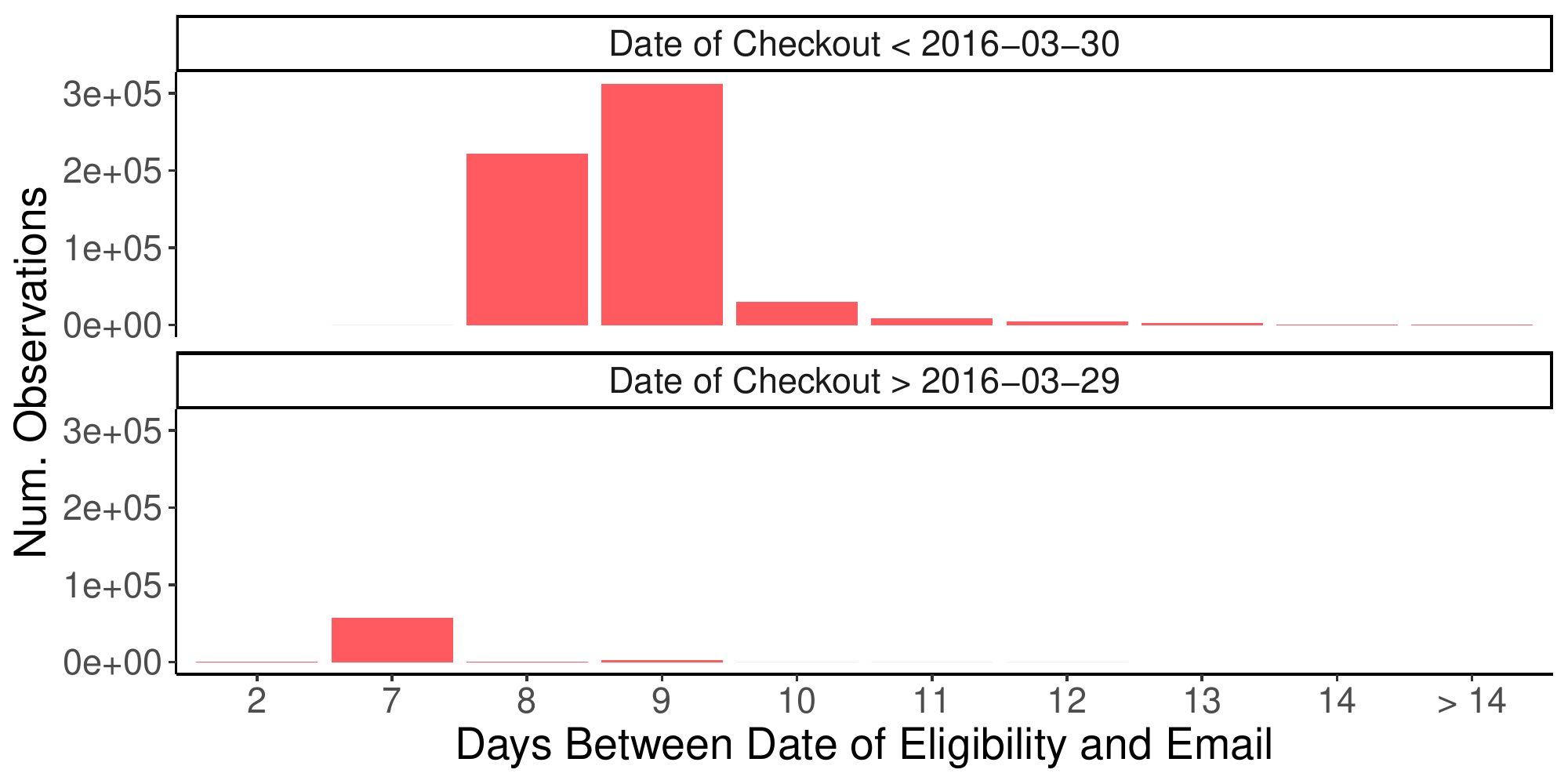}
	\caption{Days Between Assigned Date and Email}
	\label{fig:daysbetweeneligibleandemail}
	\footnotesize \emph{Notes:} \raggedright This figure plots the histogram of days between email and the assigned checkout used by the email dispatch system. Note that no email was logged for \sharemissingemail{}\% of observations, either due to missing logging or email dispatch errors. 
\end{figure}

\begin{figure}
	\includegraphics[width=\textwidth]{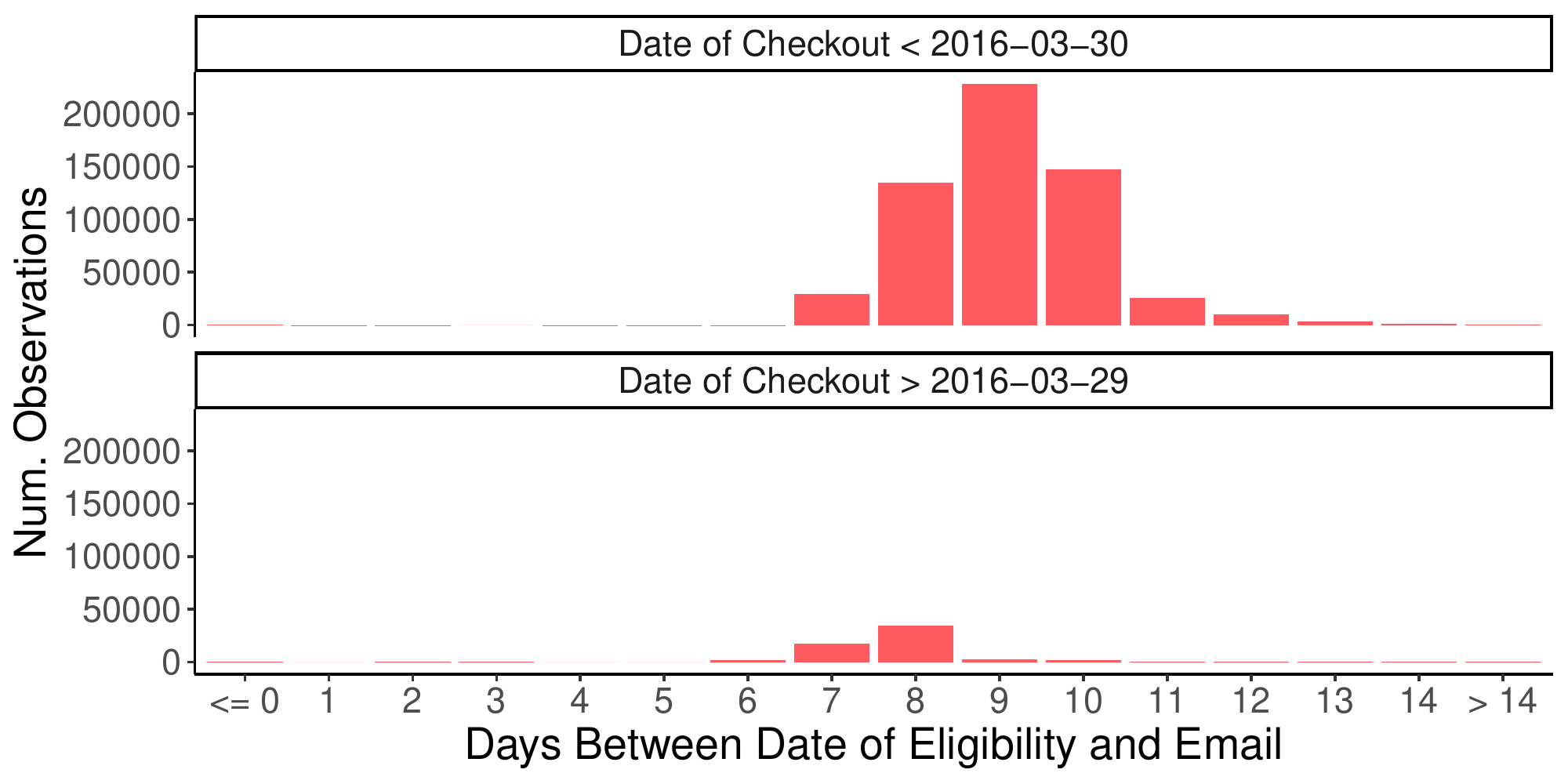}
	\caption{Days Between Realized Checkout and Email}
	\label{fig:daysbetweencheckoutandemail}
	\footnotesize \emph{Notes:} \raggedright This figure plots the histogram of days between email and the realized checkout of the focal transaction. Note that no email was logged for \sharemissingemail{}\% of observations, either due to missing logging or email dispatch errors. 
\end{figure}

\autoref{fig:daysbetweeneligibleandemail} displays histograms of days between when a listing was assigned to be reviewed by the review system and the date of the email. We can see that prior to April of 2016, the vast majority of emails were sent either 8 or 9 days after checkout. After March of 2016, most emails were sent 7 days after checkout. 
\autoref{fig:daysbetweencheckoutandemail} plots similar figures where instead the time between the true checkout (accounting for cancellations) and the email is plotted. We see that the days are more dispersed but that the pattern of time to email is similar.

We also measure differences in the days between the checkout and email across the treatment and control groups. On average, emails sent in the treatment arrived \minsbtwndiff{} minutes later after checkout than emails in the control group. This difference is statistically significant although not economically meaningful.  We do not know the exact reason for this difference but suspect it has something to do with the way in which the email dispatch system batched emails. In practice, the treatment can only affect outcomes through inducing additional reviews, and this will occurs even if emails arrive at slightly different times between the treatment and control group.

\subsection{Effects on Textual Reviews}\label{app:text}

In order to measure changes in the textual content of the reviews left by guests, we estimated the sentiment of each review in our sample using DistilBERT \citep{sanh2019distilbert}, which is a lightweight version of BERT, a widely used language model \citep{devlin2018bert}. At a high-level, BERT is a model that first pre-trains embedding-based language representations using both the left and right context around words. These pre-trained representations can then be fine-tuned to create models for a wide variety of natural language processing tasks, such as question answering, language inference, and sentiment analysis. We estimate the sentiment of each review in our sample using the default distilBERT sentiment transformer provided by Huggingface \citep{wolf2020transformers}, which has been fine-tuned on Version 2 of the Stanford Sentiment Treebank \citep{socher2013recursive}, a sentiment analysis training set consisting of 11,855 sentences taken from movie reviews.

We find that treated reviews are less likely to have text classified as positive. In particular, \meanposctr{}\% of reviews in the control group and \meanpostrt{}\% of reviews in the treatment are classified as positive ($p < 3.9 \times 10^{-9}$). Treated reviews are also \perlendif{}\% shorter in length than control reviews. 

To investigate whether the changes in review text are consistent with the changes in the star ratings, we regress the text sentiment on indicators for the treatment and the star rating. In particular, we run a regression of the following form:
\begin{equation}\label{eqn:textstar}
	text\_pos_{l} = \beta_{0} + \beta_{1} T_{l} + \gamma_{r} + \epsilon_{l}
\end{equation}
where $text\_pos_{l}$ is an indicator for whether review text is classified as positive, $T_{l}$ is a treatment indicator, and $\gamma_{r}$ are star rating fixed effects.

\autoref{tab:textcondrating} displays the results of \autoref{eqn:textstar}. Column 1 shows that review text in the treatment is less likely to be classified as positive. Column 2 shows that conditional on star ratings, review text is similar between treatment and control listings. Column 2 also shows that star ratings are highly correlated with text sentiment. Reviews with a one star rating have positive text less than 10\% of the time while reviews with a five star rating have positive text more than 99\% of the time. As a result, we conclude that incentivized reviews differ from regular reviews in similar ways whether measured by text or by rating. 

\begin{table}
	\caption{Text Sentiment Conditional on Rating}
	\label{tab:textcondrating}
	\centering
	\footnotesize
		\begin{tabular}{lcc}
\toprule
&\multicolumn{2}{c}{Text Sentiment Positive}\\
&(1) & (2)\\
\midrule Constant & 0.9405$^{***}$ & 0.0951$^{***}$\\
  &(0.0010) & (0.0062)\\
Treatment & -0.0080$^{***}$ & -0.0016\\
  &(0.0013) & (0.0010)\\
2 Stars &    & 0.1677$^{***}$\\
  &   & (0.0106)\\
3 Stars &    & 0.6122$^{***}$\\
  &   & (0.0079)\\
4 Stars &    & 0.8602$^{***}$\\
  &   & (0.0062)\\
5 Stars &    & 0.8979$^{***}$\\
  &   & (0.0062)\\
 &  & \\
R$^2$ & 0.00026&0.42832\\
Observations & 135,670&135,670\\
\bottomrule
\end{tabular}

		\vspace{.1in}
		\footnotesize \emph{Notes:} \raggedright This table plots regressions results where the outcome is the classified sentiment of the review text and the controls include a treatment indicator and star rating fixed effects.
\end{table}

\subsection{Additional Analysis of the Effects of Treatment on Listing Outcomes}\label{app:additionalresults}

In this section, we conduct additional analysis of our experimental results. In particular, we investigate whether adding controls substantially effects the precision of our estimates and whether the effects of the treatment on reservations come from the intensive or the extensive margin.

In \autoref{tab:listingoutcomescontrols} we display the results of the intent to treat regressions with a 120 day time horizon, with control variables for listing, guest, and focal transaction characteristics. In particular, we controls for room type, capacity, bedrooms, prior nights, prior bookings, trips in process, number of listings managed by the host, main photo size, number of photos, guest gender, whether the guest is a host, guest prior nights, guest prior reviews submitted, guest prior five star reviews submitted are included along with checkout week and zip code fixed effects. With these covariates, we find effects on views and reservations, but not on nights and booking value. This mirrors the results without control variables. 

Next, we consider whether the effects on reservations come from the intensive or the extensive margin. Induced reviews may help some listings who would've otherwise failed on the platform or the may hurt some listings with a negative review. For both of these cases, we would expect to see an effect on the extensive margin, i.e. whether a listing gets subsequent reservations. On the other hand, if induced reviews affect the types and frequency of subsequent transactions, then this effect may be felt on the intensive margin.

In \autoref{tab:intensivemargreg}, we estimate separate regressions where the outcome is whether a listing has a reservation at all, and how many reservations a listing has conditional on receiving at least one subsequent booking within a set number of days after the focal transaction. Columns (1), (3), and (5) show estimates for the extensive margin and fail to find economically or statistically meaningful effects. Columns (2), (4), and (6) display results for the intensive margin. There are larger in percentage terms and statistically significant effects for the 2 month and 4 month horizon. At the 12 month horizon, results are similar in levels but standard errors are much wider.  

\subsection{Why Do Reviews Affect Views?}\label{app:views}

In this section, we investigate the mechanisms behind the fact that the treatment group has more views than the control group. There are two main hypotheses for why this effect exists. The first is that searchers can see the number of reviews on the search page, and are induced to click on the listing because of this information. The second is that the ranking algorithm may take into account reviews and display reviewed listings higher.

\begin{figure}
	\caption{Effect on Search Rank and Views from Search}
	\begin{center}\label{fig:viewsdecomp}
	\includegraphics[width = 6in]{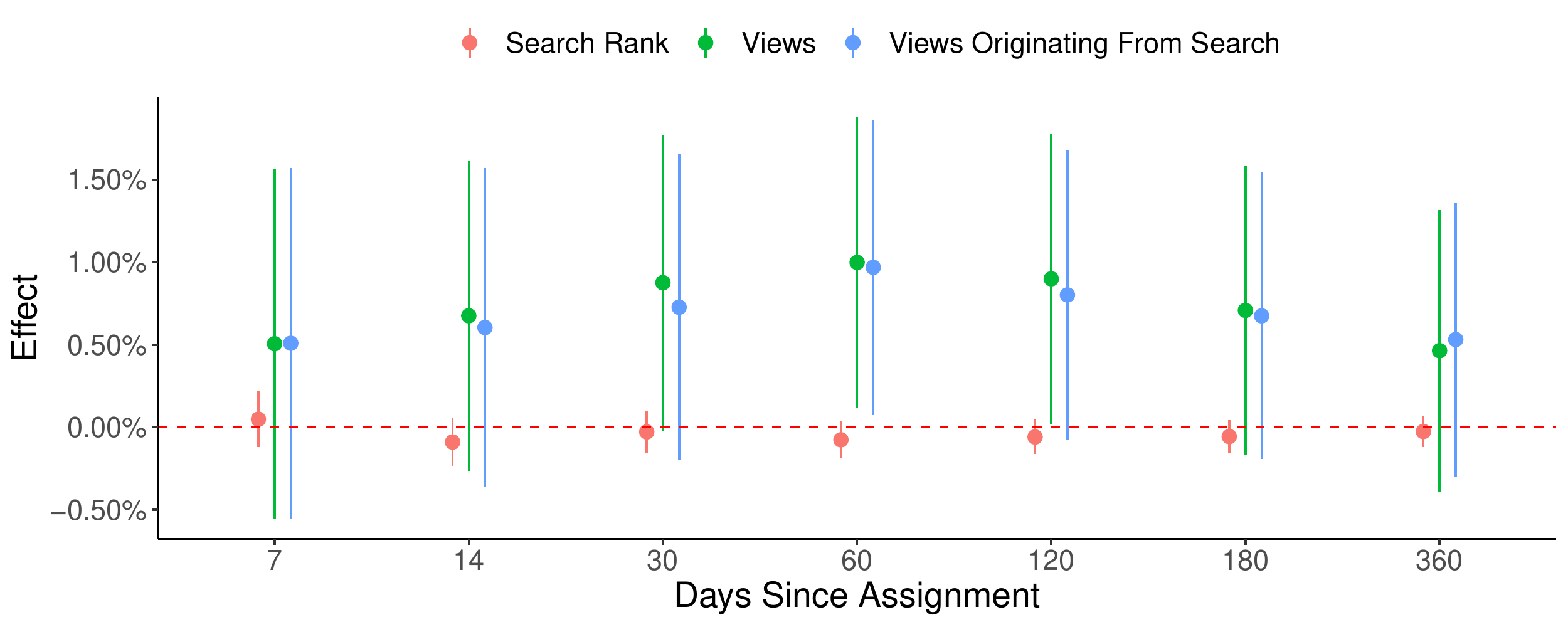}
	\end{center}\footnotesize \emph{Notes:} \raggedright This figure plots observational estimates and 95\% confidence intervals of the effect of the incentivized review email on views of a listing's page, views originating from search, and the search rank from which the views arrived. 
\end{figure}

To disentangle this, we measure whether a view originated from search and the search ranking of the listing on a search page prior to a click onto a listings.  \autoref{fig:viewsdecomp} shows the effects of the treatment on views originating from a search. These effect on views from search is similar to the effects on overall views. We also measure the effect on the originating search rank. We find a precise zero effect on the search ranking of listings prior to a view. 

These results show that views to a listing increase in the treatment but the search ranking does not change. We conclude that the presence of information about reviews in search results matters. Searchers see that treated listings have more reviews and this induces them to click on their their listing page to view more information.

\clearpage
\section{Additional Figures and Tables}
\begin{figure}[!htpb]
	\includegraphics[width=\textwidth]{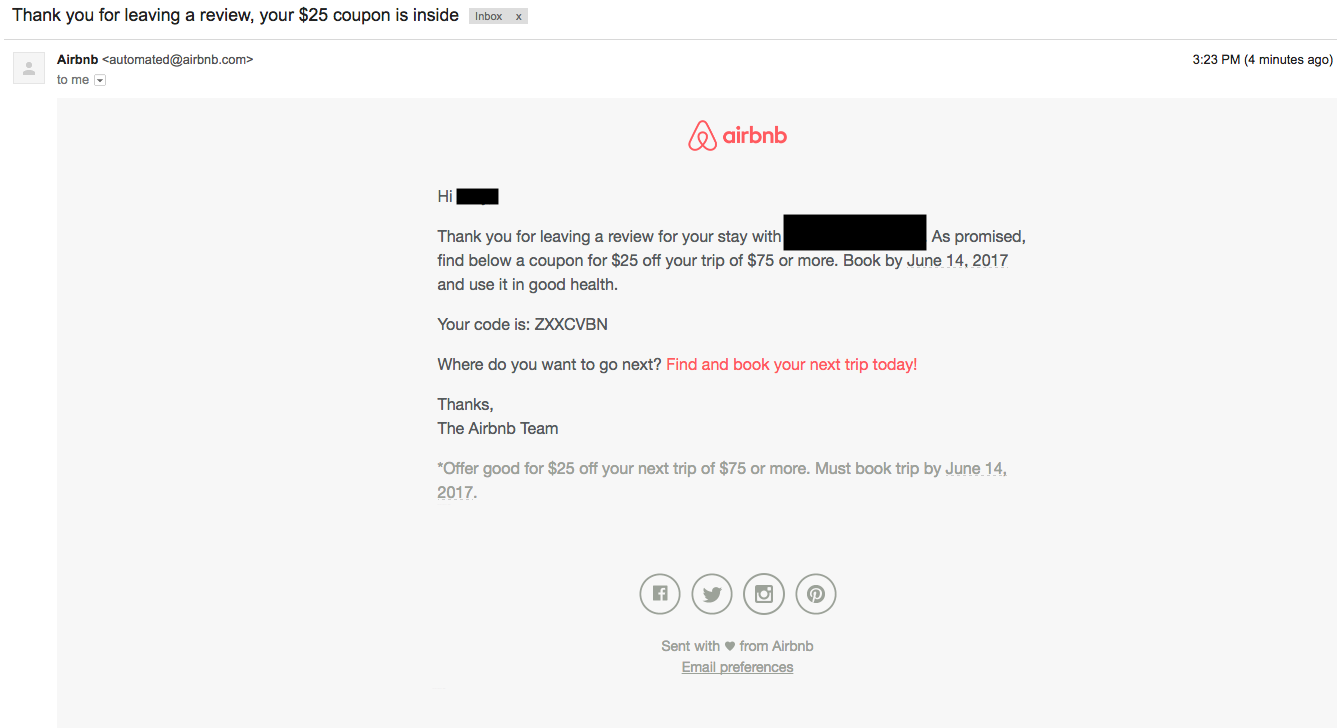}
	\caption{Email Sent After Incentivized Review}
	\label{fig:incent_email_coupon_after}
	\footnotesize \emph{Notes:} \raggedright Displays the email sent to guests who had stayed in treatment listings that had not yet received a review on Airbnb after a certain number of days, issuing them a coupon after leaving a review.
\end{figure}

\begin{figure}
	\includegraphics[width=\textwidth]{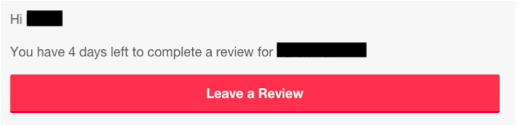}
	\caption{Email Sent to the Control Group}
	\label{fig:emailconotrol}
	\footnotesize \emph{Notes:} \raggedright Displays the email sent to guests who had stayed in control listings that had not yet received a review on Airbnb after a certain number of days.
\end{figure}

\begin{figure}[!htpb]
	\includegraphics[width=\textwidth]{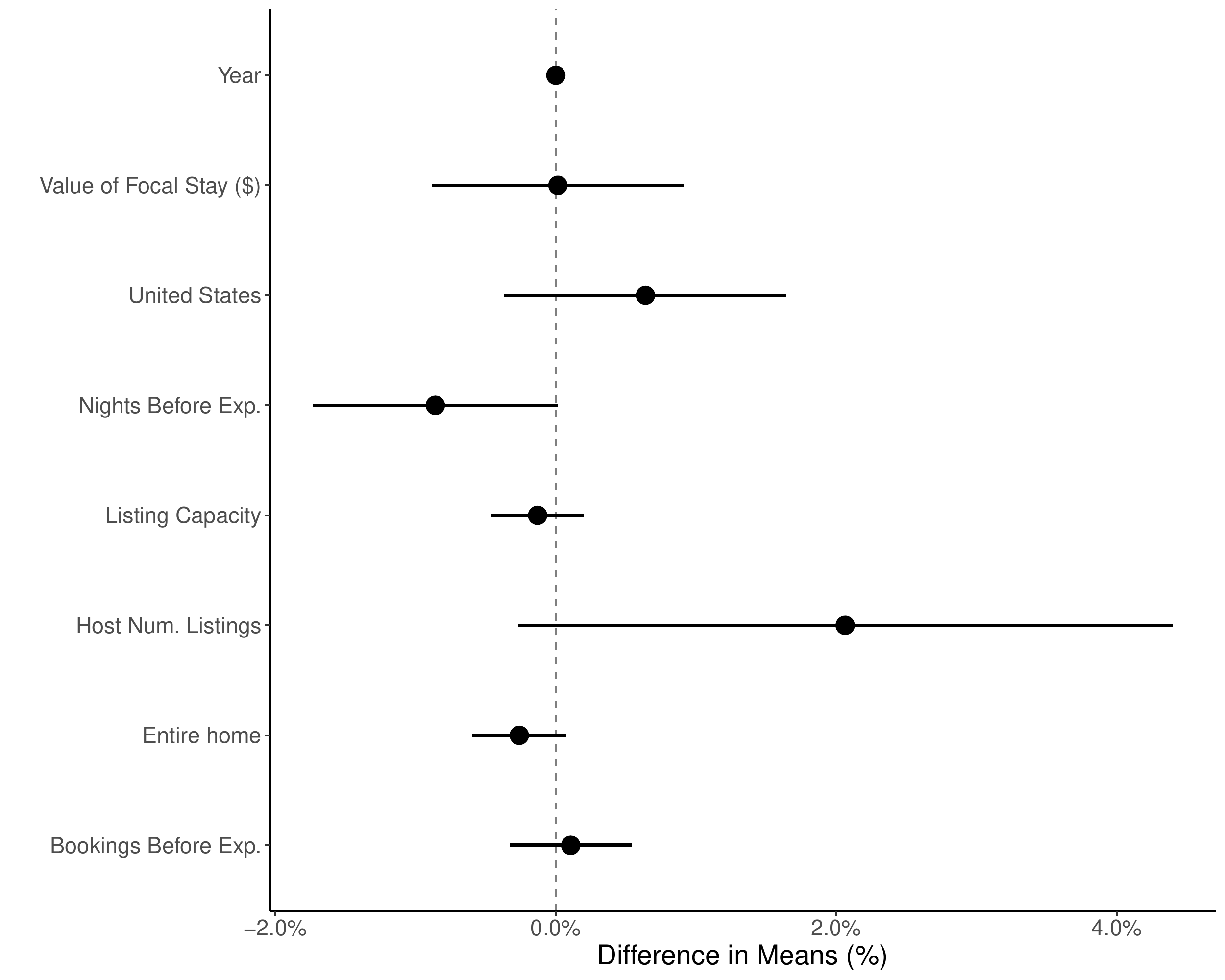}
	\caption{Balance Assessment for Experiment}
	\label{fig:balance}
	\footnotesize \emph{Notes:} \raggedright This plot displays the difference in means between the treatment and control groups for pre-treatment covariates and the days between checkout and email date. No differences were statistically significant in the pre-treatment covariates (year of stay, dollar value of transaction, whether the listing was in the United States, the number of nights the listing hosted for prior to the experimental assignment, the listing capacity, the number of listings by the host, whether the listing was an entire property and the number of bookings prior to the experiment). The days between (coupon / reminder) email and checkout is measured for a subset of listings and exhibits a slight and statistically significant difference between treatment and control.
\end{figure}

\begin{figure}[!htpb]
	\includegraphics[width=\textwidth]{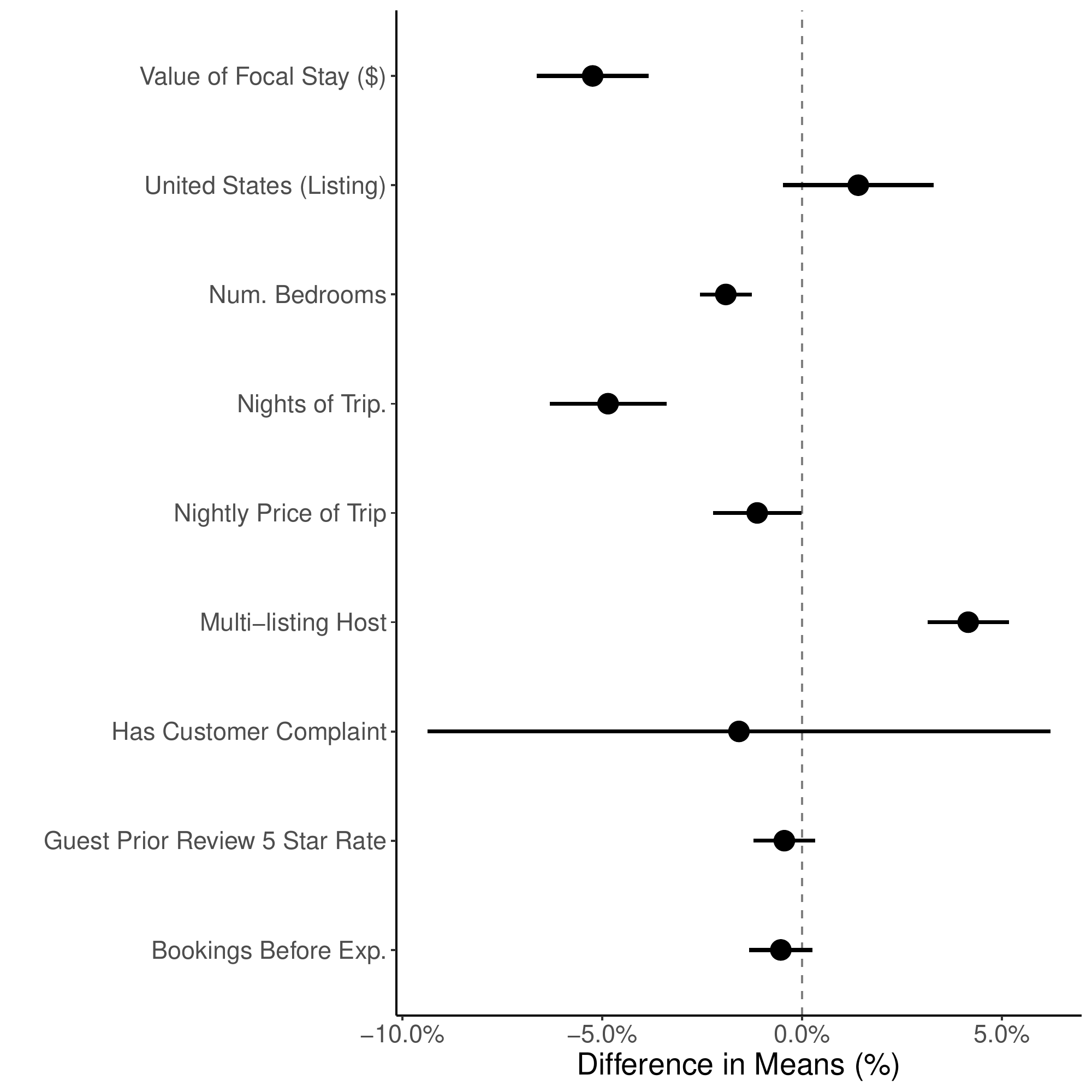}
	\caption{Differences in Characteristics of Reviewed Transactions \\ Treatment vs Control}
	\label{fig:typesofrevs}
	\footnotesize \emph{Notes:} \raggedright This plot displays the difference in means between the treatment and control groups for trip characteristics.
\end{figure}

\begin{figure}[!htpb]
	\begin{center}
	\includegraphics[width=\textwidth]{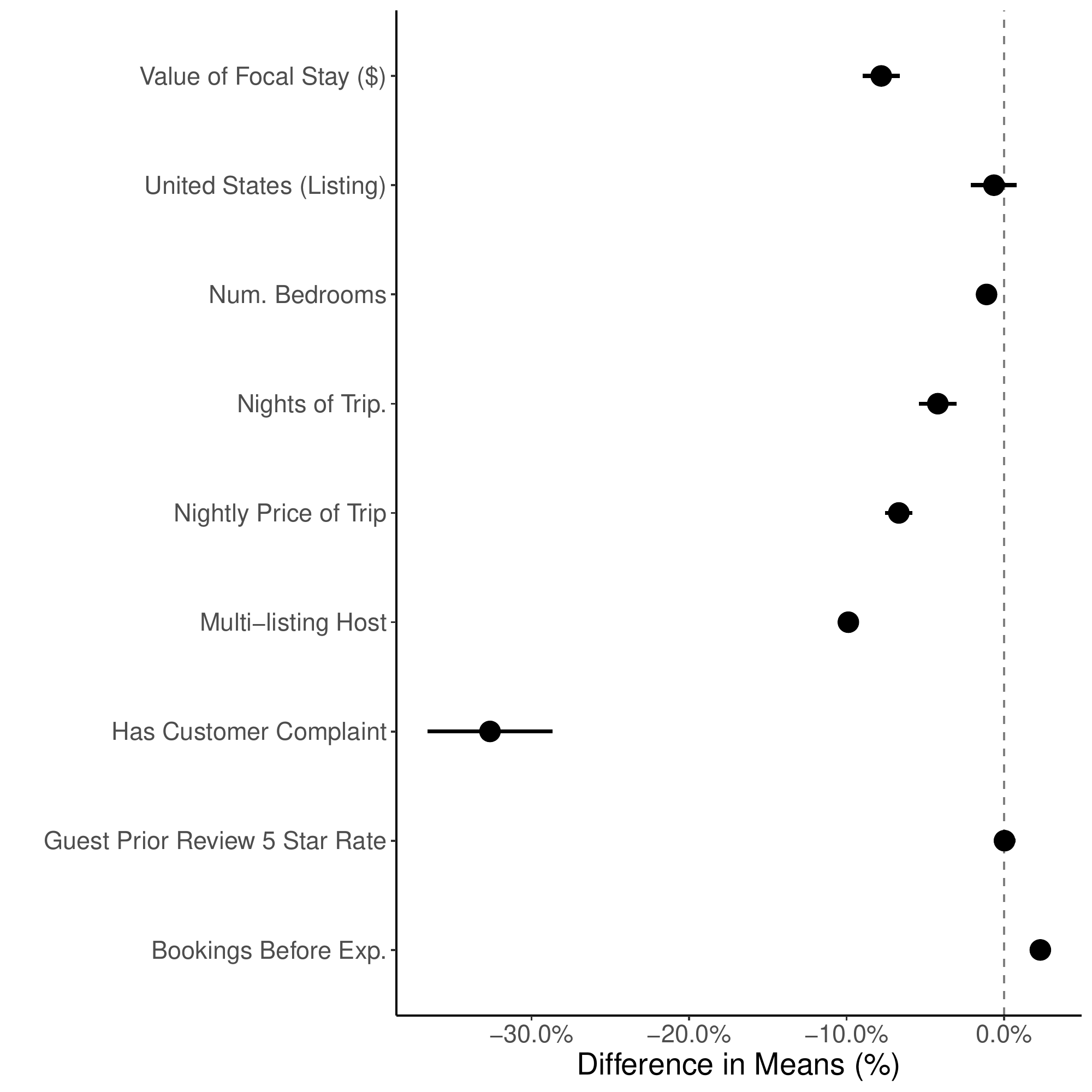}
	\caption{Differences in Characteristics of Transactions \\ Reviewed vs Non-Reviewed}
	\label{fig:typesofrevsnoncomp}	
	\end{center}
	\footnotesize \emph{Notes:} \raggedright This plot displays the difference in means between reviewed and non-reviewed transactions in the treatment versus the control groups.
\end{figure}

\clearpage

\begin{figure}
	\caption{Observational Estimate of Effect of Review}
	\begin{center}\label{fig:corrrating}
	\includegraphics[width = 6in]{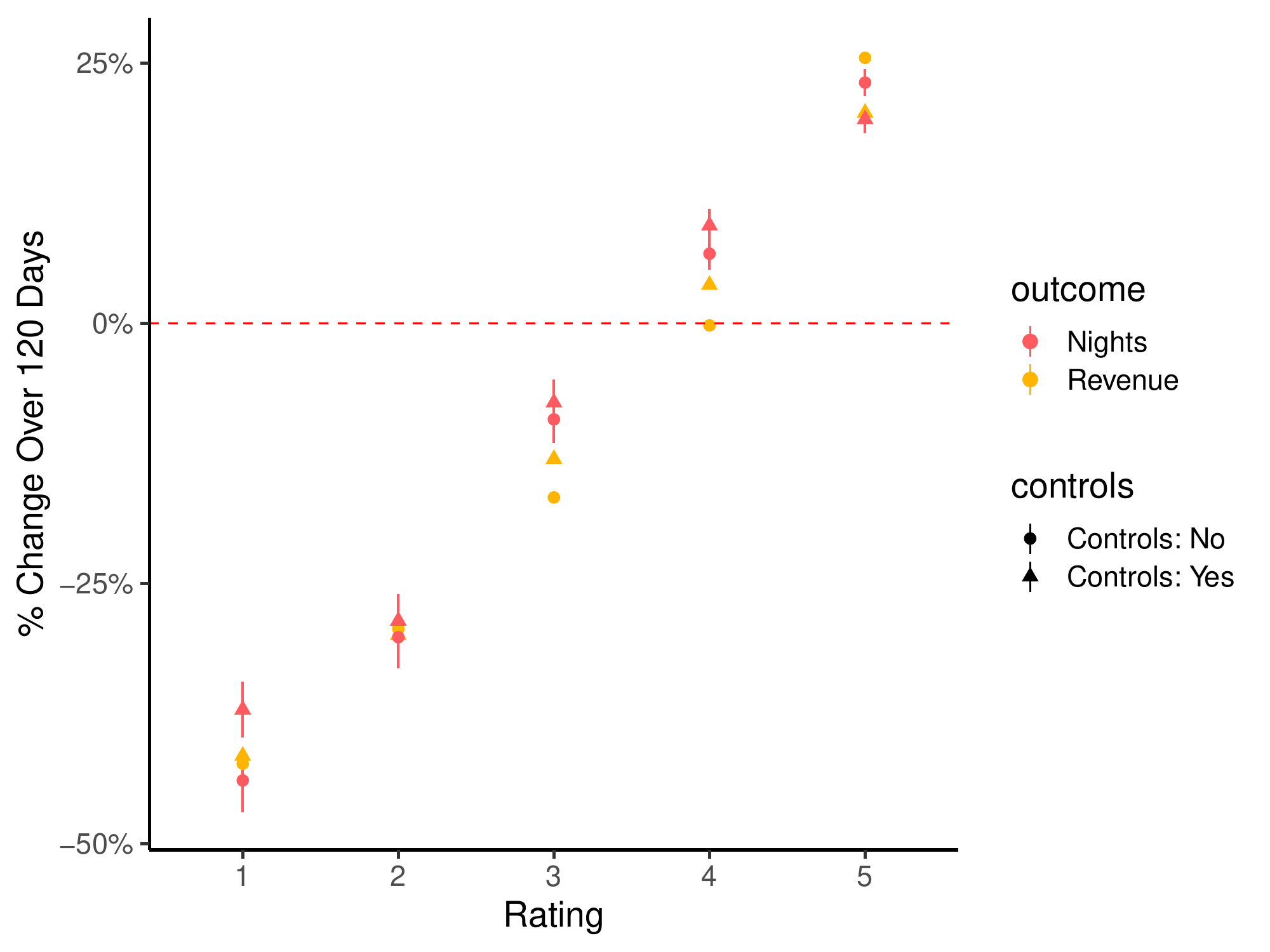}
	\end{center}\footnotesize \emph{Notes:} \raggedright This figure plots observational estimates and 95\% confidence intervals of the effect of a first review with a given star rating (1 - 5) on subsequent nights and revenue. Estimates without controls are represented by circles while estimates with controls are represented by triangles. Controls for room type, capacity, bedrooms, prior nights, prior bookings, trips in process, number of listings managed by the host, main photo size, number of photos, guest gender, whether the guest is a host, guest prior nights, guest prior reviews submitted, guest prior five star reviews submitted are included along with checkout week and market fixed effects. 
\end{figure}


\begin{table}
	\caption{Effects of Treatment on Demand with Covariates}
	\label{tab:listingoutcomescontrols}
	\centering
	\footnotesize
		\begin{tabular}{lcccc}
\toprule
&(1) & (2) & (3) & (4)\\
&Views&Reservations&Nights&Booking Value\\
\midrule Assigned to Treat. & 8.876$^{**}$ & 0.0431$^{**}$ & 0.0182 & 1.614\\
  &(3.477) & (0.0173) & (0.0689) & (9.450)\\
 &  & & & \\
R$^2$ & 0.31250&0.34640&0.30278&0.32735\\
Observations & 649,266&649,266&649,266&649,266\\
Controls & $\checkmark$ & $\checkmark$ & $\checkmark$ & $\checkmark$\\
 &   &   &   &  \\
Checkout Week FE & $\checkmark$ & $\checkmark$ & $\checkmark$ & $\checkmark$\\
Zip Code FE & $\checkmark$ & $\checkmark$ & $\checkmark$ & $\checkmark$\\
\bottomrule
\end{tabular}

		\vspace{.1in}
		\footnotesize  \raggedright \emph{Notes:} This table displays linear regression estimates measuring the effects of the treatment (the guest receiving an email with an offer of a coupon in exchange for a review) on measures of demand. `Listing Views' refers the number of times the listing's page was viewed, `Reservations' refers to the number of transactions, `Nights' refers to the number of nights that the listing was occupied, and `Booking Value' is the amount paid by guests for transactions involving this listing. All four metrics are calculated for outcomes up to 120 days since the assignment end of the focal transaction. The focal transaction is the first transaction for a listing for which it was eligible for the experiment. Controls for room type, capacity, bedrooms, prior nights, prior bookings, trips in process, number of listings managed by the host, main photo size, number of photos, guest gender, whether the guest is a host, guest prior nights, guest prior reviews submitted, guest prior five star reviews submitted are included along with checkout week and zip code fixed effects. 
	\end{table}

\begin{table}
	\caption{Intensive Margin Regression}
	\label{tab:intensivemargreg}
	\centering	
	\footnotesize 
		
\begin{tabular}{lcccccc}
\toprule
 & Has Res. & Num. Res. & Has Res. & Num. Res. & Has Res. & Num. Res.\\
 & \multicolumn{2}{c}{\textbf{2 Months After}} & \multicolumn{2}{c}{\textbf{4 Months After}} & \multicolumn{2}{c}{\textbf{12 Months After}} \\ 
 & (1) & (2) & (3) & (4) & (5) & (6)\\
\midrule Constant & 0.5081$^{***}$ & 4.191$^{***}$ & 0.5846$^{***}$ & 6.270$^{***}$ & 0.7040$^{***}$ & 12.34$^{***}$\\
  & (0.0009) & (0.0121) & (0.0009) & (0.0186) & (0.0008) & (0.0383)\\
Assigned to Treatment & 0.0016 & 0.0341$^{**}$ & 0.0012 & 0.0586$^{**}$ & 0.0006 & 0.0621\\
  & (0.0012) & (0.0171) & (0.0012) & (0.0262) & (0.0011) & (0.0541)\\
Sub-sample & All & Cond. on Res. & All & Cond. on Res. & All & Cond. on Res.\\
  &   &   &   &   &   &  \\
Observations & 654,595 & 333,125 & 654,595 & 383,029 & 654,595 & 461,008\\
\bottomrule
\end{tabular}

		\vspace{.1in}
		\footnotesize  \raggedright \emph{Notes:} This table displays OLS regression estimates measuring the effects of being assigned to treatment on intensive and extensive margin outcomes. `Has Res.' is a binary indicator for whether the listing has received a reservation after being assigned to the experiment and within a given time period (60, 120, and 360 days respectively). `Num. Res.' is the number of reservations after being assigned to the experiment, for the subsample of observations that have at least one reservation in the time period after the experiment assignment. Robust standard errors are reported. 
	\end{table}

	\begin{table}
	\caption{Effects of Treatment on Transaction Quality - With Covariates}
	\label{tab:matchqctr}
	\centering
	\footnotesize
		
\begin{tabular}{lccc}
\toprule
 & Complaint & Reviewed & Star Rating\\
 & (1) & (2) & (3)\\
\midrule Treatment & $-4.26\times 10^{-5}$ & 0.0047$^{***}$ & -0.0050$^{**}$\\
  & (0.0001) & (0.0008) & (0.0019)\\
  &   &   &  \\
R$^2$ & 0.00373 & 0.03499 & 0.02976\\
Observations & 2,431,085 & 2,431,085 & 1,579,132\\
  &   &   &  \\
Controls & Yes & Yes & Yes\\
Guest Region FE & $\checkmark$ & $\checkmark$ & $\checkmark$\\
Checkout Week FE & $\checkmark$ & $\checkmark$ & $\checkmark$\\
Num. Nights FE & $\checkmark$ & $\checkmark$ & $\checkmark$\\
Num. Guests FE & $\checkmark$ & $\checkmark$ & $\checkmark$\\
\bottomrule
\end{tabular}

		\vspace{.1in}
		\footnotesize \emph{Notes:} \raggedright  This table displays regressions measuring the effects of the treatment (the guest receiving an email with an offer of a coupon in exchange for a review) on measures of transaction quality. The set of transactions considered for this regression includes all transactions for which the checkout date was between the checkout date of the focal transaction and 360 days after. `Complaint' refers to whether a guest submitted a customer service complaint to Airbnb, `Reviewed' refers to whether the guest submitted a review, `Star Rating' refers to the star rating of any submitted reviews. Control variables include the log of transaction amount, the number of times the guest has reviewed and reviewed with a five star ratings in the past, the prior nights of the guest, whether the guest has an about description, and guest age on the platform
	\end{table}

	\begin{table}
		\caption{Heterogeneity Analysis - By Covariate}
		\label{tab:heterogsamp}
		\centering
		\footnotesize
			
\begin{tabular}{lcccccc}
\toprule
 & \multicolumn{6}{c}{Reservations Within 120 Days}\\
 & (1) & (2) & (3) & (4) & (5) & (6)\\
\midrule (Intercept) & 2.315$^{***}$ & 3.675$^{***}$ & 3.668$^{***}$ & 3.569$^{***}$ & 6.216$^{***}$ & 2.097$^{***}$\\
  & (0.0105) & (0.0124) & (0.0183) & (0.0237) & (0.0614) & (0.0110)\\
Treatment & 0.0356$^{*}$ & 0.0373$^{*}$ & 0.1269 & 0.0376$^{*}$ & 0.0397$^{*}$ & 0.0359$^{*}$\\
  & (0.0169) & (0.0175) & (0.4053) & (0.0175) & (0.0175) & (0.0169)\\
Age $<$ 30 Days & 3.592$^{***}$ &    &    &    &    &   \\
  & (0.0280) &    &    &    &    &   \\
Treatment $\times$ Age $<$ 30 Days (Demeaned) & 0.0491 &    &    &    &    &   \\
  & (0.0396) &    &    &    &    &   \\
Superhost &    & 2.581$^{***}$ &    &    &    &   \\
  &    & (0.1382) &    &    &    &   \\
Treatment $\times$ Superhost (Demeaned) &    & 0.0523 &    &    &    &   \\
  &    & (0.1922) &    &    &    &   \\
Multi-listing Host &    &    & 0.0842$^{***}$ &    &    &   \\
  &    &    & (0.0248) &    &    &   \\
Treatment $\times$ Multi-listing Host (Demeaned) &    &    & 0.0077 &    &    &   \\
  &    &    & (0.0351) &    &    &   \\
Female Host &    &    &    & 0.0481 &    &   \\
  &    &    &    & (0.0303) &    &   \\
Male Host &    &    &    & 0.3783$^{***}$ &    &   \\
  &    &    &    & (0.0326) &    &   \\
Treatment $\times$ Female Host (Demeaned) &    &    &    & 0.0336 &    &   \\
  &    &    &    & (0.0428) &    &   \\
Treatment $\times$ Male Host (Demeaned) &    &    &    & 0.0223 &    &   \\
  &    &    &    & (0.0461) &    &   \\
Log Price &    &    &    &    & -0.5649$^{***}$ &   \\
  &    &    &    &    & (0.0130) &   \\
Treatment $\times$ Log Price (Demeaned) &    &    &    &    & -0.0427$^{*}$ &   \\
  &    &    &    &    & (0.0185) &   \\
$>$ 1 Booking Prior &    &    &    &    &    & 3.551$^{***}$\\
  &    &    &    &    &    & (0.0253)\\
Treatment $\times$ $>$ 1 Booking Prior &    &    &    &    &    & 0.0330\\
  &    &    &    &    &    & (0.0357)\\
  &   &   &   &   &   &  \\
Observations & 640,893 & 640,786 & 640,936 & 640,854 & 640,936 & 640,936\\
R$^2$ & 0.06324 & 0.00223 & $4.59\times 10^{-5}$ & 0.00058 & 0.00492 & 0.06416\\
\bottomrule
\end{tabular}

			\vspace{.1in}
			\footnotesize  \raggedright \emph{Notes:} This table displays estimates of heterogenous treatment effects on reservation within 120 days of the focal stay. `Treatment' refers to the guest of the focal transaction being sent an email offering a coupon. `Age $<$ 30 Days' refers to a listing being after for fewer than 30 days prior to the focal checkout. `Multi-listing host' refers to a host having more than 1 active listing. In the gender heterogeneity regressions, the omitted category no gender information. `Log Price' is the log of the nightly price paid by the guest (inclusive of fees). `$>1$ Booking Prior' takes the value of 1 if the listing had more than 1 booking prior to checkout of the focal stay.
		\end{table}

	\begin{table}
		\caption{Change in Listing Characteristics Over a Year}
		\label{tab:moralhazard}
		\centering
		\footnotesize
			\begin{tabular}{lcc}
\toprule
&Num. Photos Changed&Description Length Changed\\
&(1) & (2)\\
\midrule (Intercept) & 0.3580$^{***}$ & 0.4324$^{***}$\\
  &(0.0008) & (0.0009)\\
Treatment & -0.0006 & 0.0011\\
  &(0.0012) & (0.0012)\\
 &  & \\
Observations & 653,907&653,907\\
\bottomrule
\end{tabular}

			\vspace{.1in}
			\footnotesize  \raggedright \emph{Notes:} This table the results of a linear regression where the outcome variable is whether the number of photos or the length of the description changed for listings between the start of the treatment and 360 days later. Fewer than 1000 observations were dropped because they could not be matched with listing photos and descriptions in the database.
		\end{table}
\end{document}